\documentstyle[preprint,aps,eqsecnum,floats,epsfig]{revtex}

\tighten     

\def\beq{\begin{equation}}
\def\eeq{\end{equation}}

\def\beqn{\begin{eqnarray}}
\def\eeqn{\end{eqnarray}}
\def\nn{\nonumber\\}
\def\dis{\displaystyle}

\def\bfig{\begin{figure}[htbp]}
\def\efig{\end{figure}}

\def\Re{\,{\mbox{Re}\,}}
\def\Im{\,{\mbox{Im}\,}}
\def\O{{\cal O}}
\def\ea{{\it et al.}}

\def\vec#1{\mbox{\protect\boldmath $#1$}}

\def\See{\vec e^{\prime *}\cdot \vec e\,}
\def\Sss{\vec s^{\prime *}\cdot \vec s\,}
\def\Ses{\vec e^{\prime *}\cdot \vec s\,}
\def\Sse{\vec s^{\prime *}\cdot \vec e\,}

\def\Sek{\vec e^{\prime *}\cdot \hat{\vec k}\,}
\def\Ske{\vec e \cdot \hat{\vec k}'\,}

\def\Skk{\hat{\vec k}'\cdot \hat{\vec k}\,}

\def\Sk{\vec\sigma \cdot \hat{\vec k} \,}
\def\kS{\vec\sigma \cdot \hat{\vec k}'\,}

\def\Ss{\vec\sigma \cdot \vec s \,}
\def\sS{\vec\sigma \cdot \vec s'\,}

\def\Vee{\vec\sigma \cdot \vec e^{\prime *} \times \vec e\,}
\def\Vss{\vec\sigma \cdot \vec s^{\prime *} \times \vec s\,}

\def\Vkk{\vec\sigma \cdot \hat{\vec k}' \times \hat{\vec k}\,}

\def\Vek{\vec\sigma \cdot \vec e^{\prime *} \times \hat{\vec k}\,}
\def\Vke{\vec\sigma \cdot \vec e \times \hat{\vec k}'\,}

\begin{document}

\preprint{\today}

\title{Low-Energy Compton Scattering of \\
  Polarized Photons on Polarized Nucleons}

\author{
   D. Babusci$^1$,
   G. Giordano$^1$,
   A.I. L'vov$^2$,
   G. Matone$^1$,
   A.M. Nathan$^3$
}

\address{ \rule{0ex}{4ex}
   $^1$ INFN-Laboratori Nazionali di Frascati, Frascati, Italy \\
   $^2$ P.N.Lebedev Physical Institute, Moscow, Russia \\
   $^3$ Department of Physics, University of Illinois at
            Urbana-Champaign, USA}

\maketitle

\begin{abstract}

The general structure of the cross section of $\gamma N$ scattering with
polarized photon and/or nucleon in initial and/or final state is
systematically described and exposed through invariant amplitudes.  A
low-energy expansion of the cross section up to and including the order
$\O(\omega^4)$ is given which involves ten structure parameters of the
nucleon (dipole, quadrupole, dispersion, and spin polarizabilities).
Their physical meaning is discussed in detail.

Using fixed-$t$ dispersion relations, predictions for these parameters 
are obtained and compared with results of chiral perturbation 
theory.  It is emphasized that Compton 
scattering experiments at large angles 
can fix the most uncertain of these structure 
parameters.  Predictions for the cross section and double-polarization 
asymmetries are given and the convergence of the expansion
is investigated.  The feasibility of the experimental 
determination of some of the structure parameters is discussed.

\end{abstract}

\newpage

\section{Introduction}

Compton scattering on the proton at low and intermediate energies 
has thus far been studied mainly with unpolarized photons.  Many 
recent data are available on 
the unpolarized differential cross section both in the region below 
pion threshold \cite{fede91,zieg92,macg95} and in 
the Delta region 
\cite{hall93,blan96,moli96,peis96,legs97}.  They 
have led to a determination of the dipole electric and magnetic 
polarizabilities of the proton, have given useful constraints on pion 
photoproduction amplitudes near the Delta peak, and have
provided sensitive 
tests for different models of Compton scattering, such as those based 
on resonance saturation \cite{ishi80,caps92,pasc95,scho96},
chiral perturbation theory \cite{bern95,hols97a}, and dispersion 
relations \cite{pfei74,guia78,akhm81,lvov81,lvov97}.

With the advent of new experimental tools such as highly-polarized 
photon beams, polarized targets,
and recoil polarimetry
\cite{LSC}, it becomes possible to study 
the very rich spin structure of Compton scattering.
In particular, many additional 
structure parameters of the nucleon, such as the spin 
\cite{lin71,levc85,ragu93,babu97} and quadrupole \cite{rade79} 
polarizabilities, could be measured 
in such new-generation experiments and 
used for testing hadron models at low energies.  
The first attempt to determine the ``backward" spin 
polarizability from unpolarized experiments 
has recently been reported \cite{legs97a}.  Therefore, 
it is timely to give a detailed description of the appropriate 
polarization observables and their
relationship to the low-energy parameters that might be measured
in such 
experiments.  That is the main purpose of the present report.

This paper is organized as follows.  In Sect. II, we introduce
the invariant Compton scattering amplitudes.  In Sect. III
we develop a general structure for the $\gamma N$ scattering
cross section with polarized photons and/or polarized nucleons
in the initial and/or final state.  In Sect. IV we do a low-energy
expansion of the invariant amplitudes and develop formulas for
the low-energy expansion of the cross section and spin observables.
In the process, we introduce and discuss the physical meaning of
the parameters (polarizabilities) which are required to 
describe the cross section up to and including the order $\O(\omega^4)$.  
In Sect. V 
we give theoretical predictions for all these polarizabilities by using 
fixed-$t$ dispersion relations and compare them with available 
predictions of chiral perturbation theory. We 
then investigate the range of validity of our low-energy expansion
and show that it is generally valid below the pion threshold.
Finally, we investigate quantitatively the dependence of 
different observables on the polarizabilities and recommend 
particularly sensitive experiments to perform.  Some of the details
related to the definitions and physical meaning of the polarizabilities
are contained in the appendices.

\section{Invariant Amplitudes}

The amplitude $T_{fi}$ for Compton scattering on the nucleon,
\beq
\gamma(k) N(p) \to \gamma'(k') N'(p'),
\eeq
is defined by
\beq
   \langle f| S-1 |i\rangle = i(2\pi)^4 \delta^4(k+p-k'-p') T_{fi}.
\eeq
Constrainted by P and T-invariance, it can be expressed in terms of six
invariant amplitudes $T_i$ as \cite{lapi61,hear62,pfei74,akhm81,lvov81}
\beqn
\label{Prange}
T_{fi}=\bar u'(p') e^{\prime * \mu}\Big\{
   &-& \frac{P'_\mu P'_\nu}{P'^2}(T_1 + \gamma \cdot K\, T_2)
          -\frac{N_\mu N_\nu}{N^2}(T_3 + \gamma \cdot K\, T_4) \nn
   &+& i\frac{P'_\mu N_\nu -P'_\nu N_\mu}{P'^2 K^2} \gamma_5 T_5
    + i\frac{P'_\mu N_\nu +P'_\nu N_\mu}{P'^2 K^2}
     \gamma_5 \gamma \cdot K\, T_6
    \Big\} e^\nu u(p).
\eeqn
where $e$ and $e'$ are the photon polarization vectors, $u$ and $u'$
are the bi-spinors of the nucleons ($\bar u u=2m$, $m$ is the nucleon
mass), and $\gamma_5={0~1\choose 1~0}$.
The orthogonal $4$-vectors $P'$, $K$, $N$ and $Q$ are defined as
\beqn
   P'_\mu=P_\mu-K_\mu\frac{P \cdot K}{K^2},
   && \qquad P=\frac12(p+p'), \qquad  K=\frac12(k'+k), \nn
   N_\mu=\epsilon_{\mu\alpha\beta\gamma}P'^\alpha Q^\beta K^\gamma,
  && \qquad Q=\frac12(p-p')=\frac12(k'-k),
\eeqn
where the antisymmetric tensor $\epsilon_{\mu\alpha\beta\gamma}$
is fixed by the condition $\epsilon_{0123}=1$.

Unfortunately, there is no accepted standard for the definitions of
the invariant amplitudes in nucleon Compton scattering.  The definitions
we use here may differ from those used in other works.
We follow the conventions of Refs.
\cite{akhm81,lvov81,lvov97}, which are related to the
so-called Hearn-Leader amplitudes $A_i^{\rm HL}$ used in Refs.
\cite{pfei74,hear62} by
\beqn
 & T_1 =  A_1^{\rm HL}, \qquad T_3 =  A_2^{\rm HL},
  \qquad T_5 =   -A_3^{\rm HL},  \nn
 & T_2 =   -A_4^{\rm HL}, \qquad T_4 =   -A_5^{\rm HL},
  \qquad T_6 =   -A_6^{\rm HL}.
\eeqn

The amplitudes $T_i$ are functions of the two variables $\nu=(s-u)/4m$
and $t$, where
\beq
 s=(k+p)^2, \quad u=(k-p')^2, \quad t=(k-k')^2
\eeq
are the usual Mandelstam variables and
and $s+u+t=2m^2$.  These functions have no kinematical singularities
but they are subject to kinematical constraints arising from the
vanishing of the denominators in the decomposition (\ref{Prange}) in cases
of forward or backward scattering. Therefore, it is useful to define the
following linear combinations \cite{lvov81,lvov97,petr81},
\beqn
\label{defA}
  A_1=\frac1t \,[T_1 + T_3 + \nu (T_2+T_4)],
 && \qquad
  A_2=\frac1t \,[2 T_5 + \nu (T_2 + T_4)], \nn
  A_3=\frac1{\eta}\,[T_1 - T_3 - \frac{t}{4\nu}(T_2-T_4)],
 && \qquad
  A_4=\frac1{\eta}\,[2 m T_6 - \frac{t}{4\nu}(T_2-T_4)], \nn
  A_5=\frac1{4\nu}\,(T_2 + T_4), \qquad
 && \qquad
  A_6=\frac1{4\nu}\,(T_2 - T_4),
\eeqn
with
\beq
   \eta=\frac 1{m^2} (m^4 - su) = 4\nu^2 + t - \frac{t^2}{4m^2} \, .
\eeq
The amplitudes $A_i(\nu,t)$ are even functions of $\nu$, they
have no kinematical singularities or constraints, and they have dimension
$m^{-3}$.

In the Lab system (the nucleon $N$ at rest) the kinematic invariants
$\nu$, $t$ and $\eta$ read:
\beq
 \nu =  \frac12 (\omega+\omega'),
 \qquad t=-2\omega\omega'(1-z),
 \qquad \eta=2\omega\omega'(1+z),
\eeq
where $\omega=k_0$, $\omega'=k'_0$
are the photon energies, $z=\cos\theta$ is the photon scattering angle, and
\beq
  \omega'=\omega+\frac{t}{2m}
  = \omega \Big[1+\frac{\omega}{m}(1-z)\Big]^{-1}.
\eeq
We will reserve the symbols $\omega$, $\omega'$, and $z$ for these
Lab-frame variables.
Note that in the Lab frame, 
$N^\mu=(0,\vec N)$, where $\vec N=(m/2)\vec
k'\times\vec k$ is orthogonal to the reaction plane.

In terms of the $A_i$, the Compton
scattering amplitude $T_{fi}$ in the Lab frame assumes the following
form:
\beqn
\label{Tlab}
T_{fi}=\frac{1}{N(t)} \Big\{
   & & 2m\See\omega\omega'\,
    \Big[ (1-\frac{t}{4m^2})(-A_1-A_3)
        -\frac{\nu^2}{m^2}A_5 - A_6 \Big] \nn
   &+& 2m\Sss\omega\omega'\,
    \Big[ (1-\frac{t}{4m^2})(A_1-A_3)
        +\frac{\nu^2}{m^2}A_5 - A_6 \Big] \nn
   &-& 2i\Vee\nu\omega\omega'(A_5+A_6) \nn
   &+& 2i\Vss\nu\omega\omega'(A_5-A_6) \nn
   &+& i \Sk\Sse\omega^2\omega'\,
    \Big[ A_2+(1-\frac{\omega'}{m})A_4+\frac{\nu}m A_5 + A_6\Big] \nn
   &-& i \kS\Ses\omega\omega'^2
    \Big[ A_2+(1+\frac{\omega }{m})A_4-\frac{\nu}m A_5 + A_6\Big] \nn
   &-& i \Sk\Ses\omega^2\omega'
    \Big[ -A_2+(1-\frac{\omega'}{m})A_4-\frac{\nu}m A_5 + A_6\Big] \nn
   &+& i \kS\Sse\omega\omega'^2
    \Big[ -A_2+(1+\frac{\omega }{m})A_4+\frac{\nu}m A_5 + A_6\Big]
      \,\Big\} ,
\eeqn
where $N(t)=(1-t/4m^2)^{1/2}$ and the two magnetic vectors $\vec s$,
$\vec s'$ are defined as:
\beq
\vec s =\hat{\vec k}  \times \vec e , \qquad
\vec s'=\hat{\vec k}' \times \vec e'.
\eeq

\section{Cross sections and asymmetries}
\subsection{General structure of the cross section}

We will consider the general structure of the cross section and
double-polarization observables in four related reactions:
\begin{mathletters}\label{4reactions}
\beqn
   \roarrow\gamma \roarrow N \to         \gamma'          N', \\
           \gamma \roarrow N \to \roarrow\gamma'          N', \\
   \roarrow\gamma          N \to         \gamma' \roarrow N', \\
           \gamma          N \to \roarrow\gamma' \roarrow N'.
\eeqn
\end{mathletters}
We start by introducing the polarization variables.

Photon polarization properties are conveniently
described in terms of the Stokes parameters $\xi_i$ ($i=1,2,3$)
\cite{jackson} which define the photon polarization matrix density
as follows \cite{landau}:
\beq\label{Stocks-def}
   \langle \, e_\alpha e_\beta^* \, \rangle =
   \frac12(1+\vec\sigma \cdot \vec\xi)_{\alpha\beta} =  \frac12
   \pmatrix{ 1+\xi_3 & \xi_1 -i\xi_2 \cr
            \xi_1 +i\xi_2 & 1-\xi_3 }_{\alpha\beta} ~ .
\eeq
Here the photon polarization vector $e_\mu$ is taken in the radiation gauge
$\vec e \cdot \hat{\vec k}=0$ and
$\alpha,\beta=1,2$ denote either of two orthogonal directions
$x_\gamma,y_\gamma$ which are in turn orthogonal to the photon
momentum direction $z_\gamma=\hat{\vec k}$.
Such a definition of $\xi_i$ is manifestly frame-dependent; nevertheless,
the quantities
\beq
  \xi_l = \sqrt{\xi_1^2 + \xi_3^2}
\eeq
and $\xi_2$ are Lorentz invariant.
They give the degree of linear and circular polarization, respectively.
Moreover, $\xi_2= \pm 1$ corresponds to the right (left) helicity state,
provided the $xyz$-frame is right-handed.
The total degree of photon polarization is given by
$\xi=\sqrt{\xi_1^2 + \xi_2^2 + \xi_3^2}\le 1 $.
The values $\xi_1$ and $\xi_3$ separately are frame dependent, although they
are still invariant with respect to boosts or rotations in the
$x_\gamma z_\gamma$-plane.
They define the angle $\varphi$ that the electric field makes with the
$x_\gamma z_\gamma$-plane:
\beq\label{Stokes-azimuth}
\cos{2 \varphi}=\frac{\xi_3}{\xi_l},\qquad
\sin{2 \varphi}=\frac{\xi_1}{\xi_l}.
\eeq
To fix the azimuthal freedom in $\xi_1$ and $\xi_3$, we first choose a
frame in which all the momenta $\vec k$, $\vec p$, $\vec k'$ and $\vec
p^{\prime}$ are coplanar. This choice is not too restrictive and
includes both the Lab and CM frames.
In such a frame and for any polarized photon,
either $\gamma$ or $\gamma'$, we take the $y_\gamma$-axis in (\ref{Stocks-def})
to lie along the direction of $\hat{\vec k} \times \hat{\vec k}'$.
Then the appropriate $z_\gamma$-axis is given by
$\hat{\vec k}$ or $\hat{\vec k}'$,
and the $x_\gamma$-axis is directed along
$(\hat{\vec k} \times \hat{\vec k}') \times \hat{\vec k}$ or
$(\hat{\vec k} \times \hat{\vec k}') \times \hat{\vec k}'$,
respectively.  The angle $\varphi$ in (\ref{Stokes-azimuth}) gives the
angle between the electric field and the reaction plane.  Thus defined, the
Stokes parameters do not depend on further specification of the frame
and are the same in the Lab or CM frame.
We will use the prime to distinguish the Stokes parameters
for initial or final photon, $\xi_i$ and $\xi_i'$.

Note that the above-defined Stokes parameters transform under
parity as
\beq\label{Stokes:P}
  \xi_1 \stackrel{P}{\to} -\xi_1,\qquad
  \xi_2 \stackrel{P}{\to} -\xi_2,\qquad
  \xi_3 \stackrel{P}{\to}  \xi_3 \, ,
\eeq
under time inversion
$(\hat{\vec k} \to -\hat{\vec k}, \ \vec e \to -\vec e^*)$ as
\beq\label{Stokes:T}
  \xi_1 \stackrel{T}{\to} -\xi_1,\qquad
  \xi_2 \stackrel{T}{\to}  \xi_2,\qquad
  \xi_3 \stackrel{T}{\to}  \xi_3 \, ,
\eeq
and under crossing 
$(\hat{\vec k} \to \hat{\vec k}', \ \vec e \to \vec e^{\prime\, *})$ as
\beq\label{Stokes:cross}
  \xi_1 \stackrel{\mathrm{cross}}{\to}  \xi_1',\qquad
  \xi_2 \stackrel{\mathrm{cross}}{\to} -\xi_2',\qquad
  \xi_3 \stackrel{\mathrm{cross}}{\to}  \xi_3'.
\eeq

The nucleon polarization matrix density is specified by a polarization
4-vector $S$ which is orthogonal to the nucleon 4-momentum $p$
\cite{landau}:
\beq
  \langle \, u(p)\bar u(p) \, \rangle=\frac12\Big(
  \gamma \cdot p + m\Big) \Big(1+\gamma_5\gamma \cdot S \Big).
\eeq
Introducing also the polarization 3-vector $\vec\zeta$ in the nucleon
rest frame, one can relate $\vec\zeta$ and $S$ through the boost
transformation,
\beq
  \vec{S}=\vec\zeta + \frac{S_0}{p_0 + m}\vec p,
  \qquad
  S_0=\frac{\vec p \cdot \vec S}{p_0}
   =\frac{\vec p \cdot \vec\zeta}{m},
\eeq
where $\sqrt{-S^2}=|\vec\zeta|\le 1$
gives the degree of nucleon polarization.
We apply the notations $S$, $\vec\zeta$
and $S'$, $\vec\zeta'$ for the initial and final nucleons, respectively.

Note that the vectors $\vec\zeta$, $\vec\zeta'$ are frame dependent and
undergo Wigner rotation around the $y$-axis when a boost
in the reaction plane is applied.
Nevertheless, $\vec\zeta$ is the same in the Lab and CM frame, although
that is
not the case for $\vec\zeta'$.

In both the Lab and CM frames, the differential cross section of
the reactions (\ref{4reactions}) reads:
\beq\label{cs-generic}
\frac{d\sigma}{d\Omega}= \Phi^2\left|T_{fi}\right|^2
    \qquad \mbox{with} \qquad \Phi = \cases{
  \dis \frac {1}{8 \pi m}\frac{\omega'}{\omega} \,,  & Lab frame, \cr
  \dis \frac {1}{8 \pi \sqrt{s}} \,, & CM frame\rule{0ex}{4ex}.  }
\eeq
Here the square of the Lorentz-invariant matrix element $T_{fi}$,
appropriately averaged and summed over polarizations,
has the same generic form in all four cases (\ref{4reactions}),
\beq
\label{TWij-def}
 \left|T_{fi}\right|^2 =
 \sum_{i=0}^{3}[W_{0i}+ K \cdot S\,W_{1i}
       + Q \cdot S\,W_{2i}+ N \cdot S\,W_{3i}]\,\xi_i \, ,
\eeq
where we set $\xi_0=1$, and for the  moment we disregard optional primes
distinguishing polarization variables for initial and final particles.
Since $S$ and $N$ are axial vectors and $\xi_1$, $\xi_2$ have odd
P-parity, some of the invariant functions $W_{ij}$ must vanish:
\beq
  W_{01}=W_{02}  =
           W_{10}=W_{13}  =
           W_{20}=W_{23}  =
  W_{31}=W_{32}  = 0.
\eeq
In terms of the remaining $W_{ij}$, the generic expression (\ref{TWij-def})
gets the following specific form for the individual cases
given in Eq.~(\ref{4reactions}):
\begin{mathletters}
\label{TWij}
\beqn
\label{TWgN}
\left|T_{fi}(\roarrow\gamma \roarrow N \to \gamma' N')\right|^2 &=&
    W_{00} + W_{03}\xi_3  + N \cdot S\,(W_{30} + W_{33}^+\xi_3)
  \nn && \qquad {}
  +  K \cdot S\,( W_{11}^+\xi_1 + W_{12}^+\xi_2)
  +  Q \cdot S\,( W_{21}^+\xi_1 + W_{22}^+\xi_2), \\[1.5ex]
\label{TWg'N}
\left|T_{fi}(\gamma \roarrow N \to \roarrow\gamma' N')\right|^2 &=&
    W_{00} + W_{03}\xi_3'  + N \cdot S\,(W_{30} + W_{33}^-\xi_3')
  \nn && \qquad {}
  +  K \cdot S\,(-W_{11}^-\xi_1' + W_{12}^-\xi_2')
  +  Q \cdot S\,( W_{21}^-\xi_1' - W_{22}^-\xi_2'), \\[1.5ex]
\label{TWgN'}
\left|T_{fi}(\roarrow\gamma N \to \gamma' \roarrow N')\right|^2 &=&
    W_{00} + W_{03}\xi_3  + N \cdot S'\,(W_{30} + W_{33}^-\xi_3)
  \nn && \qquad {}
  + K \cdot S'\,( W_{11}^-\xi_1 + W_{12}^-\xi_2)
  + Q \cdot S'\,( W_{21}^-\xi_1 + W_{22}^-\xi_2), \\[1.5ex]
\label{TWg'N'}
\left|T_{fi}(\gamma N \to \roarrow\gamma' \roarrow N')\right|^2 &=&
    W_{00} + W_{03}\xi_3'  + N \cdot S'\,(W_{30} + W_{33}^+\xi_3')
  \nn && \qquad {}
  + K \cdot S'\,(-W_{11}^+\xi_1' + W_{12}^+\xi_2')
  + Q \cdot S'\,( W_{21}^+\xi_1' - W_{22}^+\xi_2').
\eeqn
\end{mathletters}
Note that the same functions $W_{ij}$ determine the cross section
in Eqs. (\ref{TWgN}) and (\ref{TWg'N'})
(as well as in Eqs. (\ref{TWg'N}) and (\ref{TWgN'}))
which are related through T-invariance;
the negative signs in (\ref{TWg'N}), (\ref{TWg'N'})
are easily found from Eq.~(\ref{Stokes:T}).
The relationship between the cases (\ref{TWgN}) and
(\ref{TWg'N}) is determined by the crossing symmetry of the
amplitude $T_{fi}$ and Eq.~(\ref{Stokes:cross}):
\beqn
  W_{00}  (\nu,t) =  W_{00}  (-\nu,t), && \qquad
  W_{03}  (\nu,t) =  W_{03}  (-\nu,t), \nn[1.5ex]
  W_{30}  (\nu,t) = -W_{30}  (-\nu,t), && \qquad
  W_{33}^-(\nu,t) = -W_{33}^+(-\nu,t), \nn[1.5ex]
  W_{11}^-(\nu,t) =  W_{11}^+(-\nu,t), && \qquad
  W_{12}^-(\nu,t) =  W_{12}^+(-\nu,t), \nn[1.5ex]
  W_{21}^-(\nu,t) =  W_{21}^+(-\nu,t), && \qquad
  W_{22}^-(\nu,t) =  W_{22}^+(-\nu,t).
\eeqn

In terms of the invariant amplitudes $T_i$ or $A_i$,
the functions $W_{ij}$ read ({\it cf.} Ref.~\cite{frol60}):
\begin{mathletters}
\label{Wij}
\beqn
   W_{00}
    &=&\frac12(4m^2-t)(|T_1|^2+|T_3|^2)
   -\frac12(s-m^2)(u-m^2)(|T_2|^2+|T_4|^2)\nn && \qquad {}
   + m(s-u)\Re(T_1 T_2^* + T_3 T_4^*) -t|T_5|^2 +(m^4-su)|T_6|^2 \nn[1.5ex]
        &=&  \frac14 (4 m^2 -t) \Big( t^2 |A_1|^2 +
   \eta^2 |A_3|^2\Big) - \frac14\Big( t^3 |A_2|^2 -
   \eta^3 |A_4|^2\Big)  \nn && {}
   -\nu^2 t\,(t + 8 \nu^2)|A_5|^2
   +\frac12 \eta\, (t^2 + 2 m^2 \eta)|A_6|^2 \nn && {}
   +\Re\Big\{2 \nu^2 t^2(A_1 + A_2)A_5^* +
   \frac12\eta^2 (4 m^2 A_3 + t A_4)A_6^*\Big\}, \\[2ex]
   W_{03}
    &=&\frac12(4m^2-t)(|T_1|^2-|T_3|^2)
   -\frac12(s-m^2)(u-m^2)(|T_2|^2-|T_4|^2)\nn && \qquad {}
   + m(s-u)\Re(T_1 T_2^* - T_3 T_4^*) \nn[1.5ex]
        &=&  \frac{\eta t}{2} \Re\Big\{ \Big( (4 m^2 - t) A_1
    + 4 \nu^2 A_5 \Big) A_3^* \,+\, 4 m^2 A_1 A_6^* \Big\}, \\[2ex]
   W_{30}
    &=& -4\Im(T_1 T_2^* + T_3 T_4^*)  \nn
        &=& - 8 \nu\Im(t A_1 A_5^* + \eta A_3 A_6^*), \\[2ex]
%
   W_{33}^{\,\pm}
    &=&4\Im\Big[-T_1 T_2^* + T_3 T_4^* \pm 2T_5 T_6^*\Big]\nn[1.5ex]
         &=& \Im \Big\{ {-} 8\nu\, \Big[ \Big(
     t A_1 - (t+4\nu^2) A_5 \Big) A_6^*  + \eta A_3 A_5^* \Big]
  \nn && {} \qquad
     \pm \frac2m (t A_2 - 4\nu^2 A_5)(\eta A_4^* + t A_6^*)\Big\}, \\[2ex]
   W_{11}^{\,\pm}
    &=& \Im\Big[(4m^2-t)(T_1 + T_3) T_6^*
     + m(s-u)(T_2 + T_4) T_6^*
     \mp t(T_2 - T_4) T_5^* \Big] \nn[1.5ex]
        &=& \Im\Big\{ \frac{t}{2m}
   \Big( (4m^2-t)A_1 + 4\nu^2 A_5\Big) (\eta A_4^* + t A_6^*)
  \nn && \qquad {}
   \pm 2\nu t \, (t A_2 - 4\nu^2 A_5) A_6^* \Big\}, \\[2ex]
   W_{12}^{\,\pm}
    &=&\Re\Big[(4m^2-t)(-T_1 + T_3) T_6^*
     +m(s-u)(-T_2 + T_4) T_6^*
     \pm t(T_2 + T_4) T_5^* \Big] \nn[1.5ex]
      &=&  \Re \Big\{ {-} \frac{\eta}{2m}
     \Big( (4m^2-t) A_3 + 4m^2 A_6 \Big) (\eta A_4^* + t A_6^*)
  \nn && \qquad {}
   \pm 2\nu t\, (t A_2 - 4\nu^2 A_5) A_5^*  \Big\}, \\[2ex]
   W_{21}^{\,\pm}
    &=&\Im\Big[4m(T_1 - T_3) T_5^*
     +(s-u)(T_2 - T_4) T_5^* \nn && \qquad {}
     \mp mt (T_2 + T_4) T_6^*
     \pm (s-u)(T_1 + T_3) T_6^*\Big] \nn[1.5ex]
      &=& 2 \Im\Big\{   {-}m\,(t A_2 - 4\nu^2 A_5)
     \Big( \eta A_3^* + (t+4\nu^2) A_6^* \Big)  \nn && \hspace{8em} {}
      \pm \nu \Big( t A_1 - (t+4\nu^2) A_5 \Big)
     (\eta A_4^* + t A_6^*) \Big\}, \\[2ex]
   W_{22}^{\,\pm}
    &=&\Re\Big[-4m(T_1 + T_3) T_5^*
     -(s-u)(T_2 + T_4) T_5^* \nn && \qquad {}
     \pm mt (T_2 - T_4) T_6^*
     \mp (s-u)(T_1 - T_3) T_6^* \Big] \nn[1.5ex]
        &=&  2 \Re\Big\{ {-}m t\,(t A_2 - 4 \nu^2 A_5) A_1^*
   \mp \nu \eta A_3\,(\eta A_4^* + t A_6^*) \Big\}.
\eeqn
\end{mathletters}

\noindent
Below the pion photoproduction threshold, the amplitudes $T_i$, $A_i$
are real, and therefore only the six structures, $W_{00},W_{03},W_{12}^\pm$
and $W_{22}^\pm$, are different from zero.

\subsection{Asymmetries}

The thirteen invariant quantities $W_{ij}$ standing in Eq.~(\ref{TWij})
intervene directly in the definition of polarization observables
which can be measured experimentally.
Not all of them are independent, since with 6 complex
invariant amplitudes there are only 11 independent observables.
In the following we shall classify them according to the number
of polarization degrees of freedom involved in the process.
We shall define some $n$-index asymmetries
$\Sigma_i$, $\Sigma_\alpha$ and $\Sigma_{i\alpha}$,
where $i = (1,2,3)$ or $(1',2',3')$ refers to the photon
Stokes parameters $\xi_i$ or $\xi_i'$, and $\alpha = 
(x_N,y_N,z_N)$ or
$(x_N',y_N',z_N')$ 
refers to the right-handed axes along which the nucleon spin $\vec\zeta$ or
$\vec\zeta'$ may be aligned.  At this point we have to relate
the products like $K \cdot S$, $Q \cdot S$, and $N \cdot S$ in
Eq.~(\ref{TWij}) to the spin vector $\vec\zeta$ (or $\vec\zeta'$), and
eventually fix a frame.  We choose the Lab frame for practical
reasons;  for 
any other frame only 
the coefficients $C^{K,Q}_{x,z,x',z'}$ introduced below have to be
recalculated.
We will indicate all necessary changes to be done for the CM frame.

Directing the $y_N$ and $y_N'$ axes along $\vec k \times \vec k'$,
we choose the $z_N$-axis along the photon beam direction $\hat{\vec k}$.
Note that since $\vec\zeta_{\rm CM}=\vec\zeta_{\rm Lab}$,
all the asymmetries obtained in this way in the Lab frame for the
reactions (\ref{4reactions}-a,b) with the polarized nucleon $N$ are
the same in the CM frame.  In the case of the polarized
final nucleon, we choose the $z_N'$ axis to lie along the nucleon recoil
momentum $\vec p'$.  Such an axis depends on whether we use the Lab
or CM frame so that we get different, though related,
asymmetries.

\begin{figure}[hbtp]
\begin{center}
\leavevmode
\epsfxsize=10cm
\epsfbox[0 150 400 350]{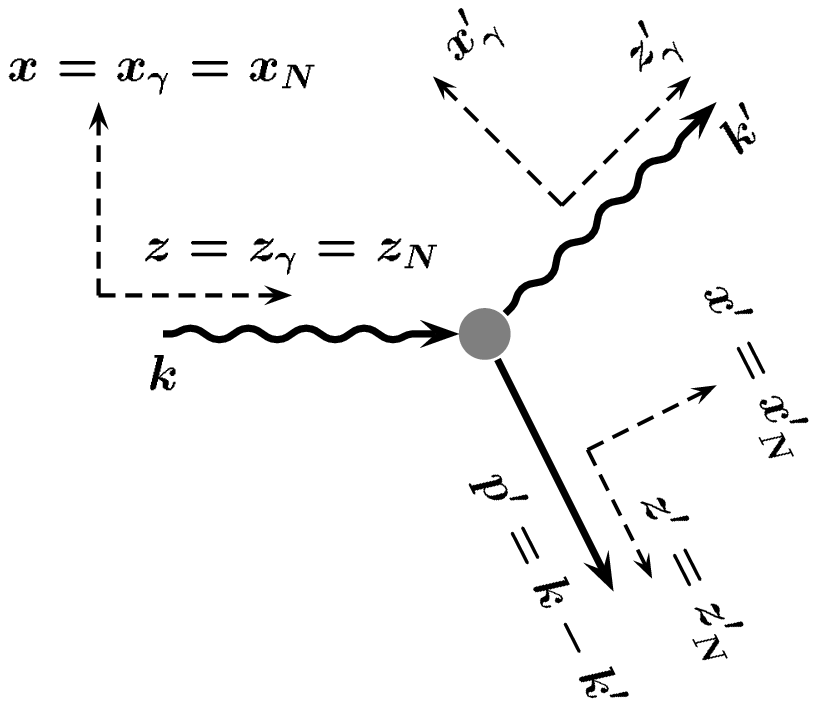}
\end{center}
\caption{Axes to project out polarizations in the Lab frame.  The various
y-axes ($y_\gamma$, $y_N$, $y_\gamma^\prime$, and
$y_N^\prime$) all point out of the plane of the figure.}
\label{fig:axes}
\end{figure}

We can further simplify the notation.  The axes
$(x_\gamma,y_\gamma,z_\gamma)$ and $(x_N,y_N,z_N)$ are identical and we
will denote them in the following by simply $(x,y,z)$. For
$(x_N',y_N',z_N')$ we will use $(x',y',z')$.  Note that, despite the
fact that the
$y$- and $y'$-axes are the same, we will keep the prime when using $y'$
because it identifies which nucleon ($N$ or $N'$) we are
referring to.  All
these axes are shown in Fig.~\ref{fig:axes}.

The set of observables is defined as follows.

\vspace{1ex}
$\bullet$  Polarization-Independent Observable ($n=0$), or simply the unpolarized
cross section,
\beq
\frac{d \bar{\sigma}}{d \Omega}= \Phi^2 W_{00} \, .
\eeq

\vspace{1ex}
$\bullet$  Single-Polarization Observables ($n=1$), of which there are
two for each of the reactions (\ref{4reactions}):

\vspace{1ex}
(i) the beam asymmetry for photons which are linearly polarized either
parallel or perpendicular to the scattering plane $(\xi_3 = \pm 1)$ and
unpolarized nucleons (target and recoil).  The same quantity gives the
degree of linear polarization ($\xi_3'$) of the photon scattered from
unpolarized nucleons:
\beq\label{Lambda}
\Sigma_3  = \Sigma_{3'}
  = \left(\frac{\sigma^\parallel - \sigma^\perp}
   {\sigma^\parallel + \sigma^\perp}\right)_{\zeta=\zeta'=0}
  = \left(\frac{\sigma^{\parallel'} - \sigma^{\perp'}}
   {\sigma^{\parallel'} + \sigma^{\perp'}}\right)_{\zeta=\zeta'=0}
  =\frac{W_{03}}{W_{00}}
\eeq
(the primed polarizations $\parallel'$ or $\perp'$
refer to the final photon state).
This asymmetry is often designated as $\Sigma$.
%

\vspace{1ex}
(ii) the target asymmetry or recoil polarization
for unpolarized photons, whereby either $N$ or $N'$ is polarized
$(\zeta_y ~\mbox{or}~ \zeta_y' = \pm 1)$
along the $\pm y=\pm y'$ directions:
\beq
\Sigma_y  = \Sigma_{y'}
  =\left(\frac{\sigma_y -\sigma_{-y}}
  {\sigma_y + \sigma_{-y}}\right)_{\xi=\xi'= 0}
  =\left(\frac{\sigma_{y'} -\sigma_{-y'}}
  {\sigma_{y'} + \sigma_{-y'}}\right)_{\xi=\xi'= 0}
  =C^{N}_{y}\,\frac{W_{30}}{W_{00}}.
\eeq
Here the coefficient
\beq
  C^{N}_y = \frac m2 \omega\omega'\sin\theta
          = \frac m4 \sqrt{-\eta t}
\eeq
(as given through both Lab and invariant quantities)
determines the scalar product $N \cdot S =C^{N}_y \zeta_y$.
It is just the negative-$y$-component of $N^\mu$.
The asymmetry $\Sigma_y=\Sigma_{y'}$ is often designated as $\cal P$,
the recoil nucleon polarization.

\vspace{1ex}
$\bullet$  Double-Polarization Observables ($n=2$), of which there are
five for each of the reactions (\ref{4reactions}):

\vspace{1ex}
(i) the beam-target asymmetry
for incoming linearly polarized photons, either
parallel or perpendicular to the scattering plane
$(\xi_3 = \pm 1)$, and target nucleon polarized perpendicular to
the scattering plane.
The same quantity can be measured
as the correlation of polarizations of the
outgoing particles $\gamma'$ and $N'$:
\beqn
\Sigma_{3y}=\Sigma_{3'y'}
   &=& \frac
  {(\sigma^\parallel - \sigma^\perp)_{ y}
  -(\sigma^\parallel - \sigma^\perp)_{-y}}
  {(\sigma^\parallel + \sigma^\perp)_{ y}
  +(\sigma^\parallel + \sigma^\perp)_{-y}} \nn
   &=& \frac
  {(\sigma^{\parallel'} - \sigma^{\perp'})_{ y'}
  -(\sigma^{\parallel'} - \sigma^{\perp'})_{-y'}}
  {(\sigma^{\parallel'} + \sigma^{\perp'})_{ y'}
  +(\sigma^{\parallel'} + \sigma^{\perp'})_{-y'}}
  =C^{N}_{y}\frac{W_{33}^+}{W_{00}}\,.
\eeqn
Another independent quantity is the correlation of the beam
polarization and the polarization of recoil nucleon, or the correlation
of the target polarization with the final photon polarization:
\beqn
\Sigma_{3y'}=\Sigma_{3'y}
   &=& \frac
  {(\sigma^\parallel - \sigma^\perp)_{ y'}
  -(\sigma^\parallel - \sigma^\perp)_{-y'}}
  {(\sigma^\parallel + \sigma^\perp)_{ y'}
  +(\sigma^\parallel + \sigma^\perp)_{-y'}} \nn
   &=& \frac
  {(\sigma^{\parallel'} - \sigma^{\perp'})_{ y}
  -(\sigma^{\parallel'} - \sigma^{\perp'})_{-y}}
  {(\sigma^{\parallel'} + \sigma^{\perp'})_{ y}
  +(\sigma^{\parallel'} + \sigma^{\perp'})_{-y}}
  =C^{N}_{y}\frac{W_{33}^-}{W_{00}}\,.
\eeqn

\vspace{1ex}
(ii) the asymmetries with circular photon
polarizations $(\xi_2 ~\mbox{or}~ \xi_2' = \pm 1)$.\\
The general expression for the asymmetry with circularly polarized photons
follows from Eq.~(\ref{TWij}). For example, for the reaction
(\ref{4reactions}a)
\beq\label{Sigma}
\frac{\sigma^R-\sigma^L}{\sigma^R+\sigma^L} =
  \frac{K \cdot S\,W_{12}^+ + Q \cdot S\,W_{22}^+}
    {W_{00}+ N \cdot S\,W_{30}}\,.
\eeq
Such a quantity survives when the nucleon spin lies in the reaction
plane.  Considering the cases with the target spin $\vec\zeta$ aligned
in $\pm x$ or $\pm z$ directions, we can introduce the beam-target
asymmetries
\beqn\label{Sigxz}
  \Sigma_{2x}
   = \frac{\sigma^R_x-\sigma^L_x}{\sigma^R_x+\sigma^L_x}
   = \frac{C^{K}_{x}W_{12}^+ + C^{Q}_{x}W_{22}^+}{W_{00}}
   \, , \nn
  \Sigma_{2z}
   = \frac{\sigma^R_z-\sigma^L_z}{\sigma^R_z+\sigma^L_z}
   = \frac{C^{K}_{z}W_{12}^+ + C^{Q}_{z}W_{22}^+}{W_{00}}
\eeqn
and the scattered photon-target asymmetries
\beqn\label{Sigxz-}
  \Sigma_{2'x}
   = \frac{\sigma^{R'}_x-\sigma^{L'}_x}{\sigma^{R'}_x+\sigma^{L'}_x}
   = \frac{C^{K}_{x}W_{12}^- - C^{Q}_{x}W_{22}^-}{W_{00}}
   \, , \nn
  \Sigma_{2'z}
   = \frac{\sigma^{R'}_z-\sigma^{L'}_z}{\sigma^{R'}_z+\sigma^{L'}_z}
   = \frac{C^{K}_{z}W_{12}^- - C^{Q}_{z}W_{22}^-}{W_{00}} \, .
\eeqn
Here the coefficients $C^{K,Q}_{x,z}$ determine the scalar products
\beq
  K \cdot S = C^{K}_x\zeta_x + C^{K}_z\zeta_z,\qquad
  Q \cdot S = C^{Q}_x\zeta_x + C^{Q}_z\zeta_z .
\eeq
In terms of Lab or invariant quantities, they read
\beqn
\label{CKx}
&&
C^{K}_x=
C^{Q}_x=-\frac12\,\omega'\sin\theta
       = - \frac{m\sqrt{-\eta t}} {2(s-m^2)},
 \nn &&
C^{K}_z=-\frac12\,(\omega + \omega'\cos\theta)
       = -\frac{s-m^2}{2m} - \frac{t\,(s+m^2)}{4m\,(s-m^2)},
 \nn &&
C^{Q}_z= \frac12\,(\omega - \omega'\cos\theta)
       = - \frac{t\,(s+m^2)}{4m\,(s-m^2)}.
\eeqn
%
%
Numerically, $C^{K,Q}_{x,z}$ are the same in the Lab and CM frames.

Another four asymmetries, which are depend linearly on those in Eqs.
(\ref{Sigxz})-(\ref{Sigxz-}), describe the correlations of the recoil
polarization with the polarization of the photons $\gamma$ or $\gamma'$:
\beqn\label{Sigxz'}
 \Sigma_{2x'}
   = \frac{\sigma^R_{x'}-\sigma^L_{x'}}{\sigma^R_{x'}+\sigma^L_{x'}}
   = \frac{C^{K}_{x'}W_{12}^- + C^{Q}_{x'}W_{22}^-}{W_{00}}
   \, , \nn
 \Sigma_{2z'}
   = \frac{\sigma^R_{z'}-\sigma^L_{z'}}{\sigma^R_{z'}+\sigma^L_{z'}}
   = \frac{C^{K}_{z'}W_{12}^- + C^{Q}_{z'}W_{22}^-}{W_{00}}
   \, , \nn
 \Sigma_{2'x'}
   =
\frac{\sigma^{R'}_{x'}-\sigma^{L'}_{x'}}{\sigma^{R'}_{x'}+\sigma^{L'}_{x'}}
   = \frac{C^{K}_{x'}W_{12}^+ - C^{Q}_{x'}W_{22}^+}{W_{00}}
   \, , \nn
 \Sigma_{2'z'}
   =
\frac{\sigma^{R'}_{z'}-\sigma^{L'}_{z'}}{\sigma^{R'}_{z'}+\sigma^{L'}_{z'}}
   = \frac{C^{K}_{z'}W_{12}^+ - C^{Q}_{z'}W_{22}^+}{W_{00}}
   \, .
\eeqn
They are given by the coefficients $C^{K,Q}_{x',z'}$ appearing in the
expansions
\beq
  K \cdot S' = C^{K}_{x'}\zeta'_{x'} + C^{K}_{z'}\zeta'_{z'},\qquad
  Q \cdot S' = C^{Q}_{x'}\zeta'_{x'} + C^{Q}_{z'}\zeta'_{z'},
\eeq
which in the Lab frame read
\beq
  C^{K}_{x'}= -\frac{\sqrt\eta}{2N(t)},\quad
  C^{Q}_{x'}= 0,\quad
  C^{K}_{z'}=  \frac{\nu\sqrt{-t}}{2mN(t)}, \quad
  C^{Q}_{z'}=  \frac{\sqrt{-t}}{2} N(t) \qquad \mbox{(Lab)},
\eeq
where $N(t)=(1-t/4m^2)^{1/2}$.
Since the $z'$-axis in the CM frame is different from that in the Lab
frame, the corresponding $C_{x',z'}^{K,Q}$ are different too. Except for the
signs, they coincide with $C^{K,Q}_{x,y}$ in Eq.~(\ref{CKx}):
\beq
  C^{K}_{x'}= -C^{Q}_{x'}= C^{K}_{x} = C^{Q}_{x},\quad
  C^{K}_{z'}= -C^{K}_{z},\quad
  C^{Q}_{z'}=  C^{Q}_{z}    \qquad \mbox{(CM)}.
\eeq

\vspace{1ex}
(iii) the asymmetries with photons linearly polarized at $\varphi = 
\pm\pi/4$ with respect to the scattering plane ($\xi_1 = \pm 1)$.
For example, in the case (\ref{4reactions}a),
\beq
\frac{\sigma^{\pi/4} - \sigma^{- \pi/4}}
 {\sigma^{\pi/4} + \sigma^{- \pi/4}}
  =\frac{K \cdot S\, W_{11}^+ + Q \cdot S\,W_{21}^+}
  {W_{00}+N \cdot S\,W_{30}}\;.
\eeq
With the same considerations discussed for Eq.~(\ref{Sigxz}),
one can define the asymmetries
\beqn\label{Sig1xz}
 \Sigma_{1x}
   = \frac{\sigma^{\pi/4}_x-\sigma^{-\pi/4}_x}
          {\sigma^{\pi/4}_x+\sigma^{-\pi/4}_x}
   = \frac{C^{K}_{x}W_{11}^+ + C^{Q}_{x}W_{21}^+}{W_{00}}
   \, , \nn
 \Sigma_{1z}
   = \frac{\sigma^{\pi/4}_z-\sigma^{-\pi/4}_z}
          {\sigma^{\pi/4}_z+\sigma^{-\pi/4}_z}
   = \frac{C^{K}_{z}W_{11}^+ + C^{Q}_{z}W_{21}^+}{W_{00}}
   \, , \nn
 \Sigma_{1'x}
   = \frac{\sigma^{\pi/4'}_x-\sigma^{-\pi/4'}_x}
          {\sigma^{\pi/4'}_x+\sigma^{-\pi/4'}_x}
   = \frac{-C^{K}_{x}W_{11}^- + C^{Q}_{x}W_{21}^-}{W_{00}}
   \, , \nn
 \Sigma_{1'z}
   = \frac{\sigma^{\pi/4'}_z-\sigma^{-\pi/4'}_z}
          {\sigma^{\pi/4'}_z+\sigma^{-\pi/4'}_z}
   = \frac{-C^{K}_{z}W_{11}^- + C^{Q}_{z}W_{21}^-}{W_{00}}
   \, ,
\eeqn
and a similar linearly dependent set
\beqn\label{Sig1xz'}
 \Sigma_{1x'}
   = \frac{\sigma^{\pi/4}_{x'}-\sigma^{-\pi/4}_{x'}}
          {\sigma^{\pi/4}_{x'}+\sigma^{-\pi/4}_{x'}}
   = \frac{C^{K}_{x'}W_{11}^- + C^{Q}_{x'}W_{21}^-}{W_{00}}
   \, , \nn
 \Sigma_{1z'}
   = \frac{\sigma^{\pi/4}_{z'}-\sigma^{-\pi/4}_{z'}}
          {\sigma^{\pi/4}_{z'}+\sigma^{-\pi/4}_{z'}}
   = \frac{C^{K}_{z'}W_{11}^- + C^{Q}_{z'}W_{21}^-}{W_{00}}
   \, , \nn
 \Sigma_{1'x'}
   = \frac{\sigma^{\pi/4'}_{x'}-\sigma^{-\pi/4'}_{x'}}
          {\sigma^{\pi/4'}_{x'}+\sigma^{-\pi/4'}_{x'}}
   = \frac{-C^{K}_{x'}W_{11}^+ + C^{Q}_{x'}W_{21}^+}{W_{00}}
   \, , \nn
 \Sigma_{1'z'}
   = \frac{\sigma^{\pi/4'}_{z'}-\sigma^{-\pi/4'}_{z'}}
          {\sigma^{\pi/4'}_{z'}+\sigma^{-\pi/4'}_{z'}}
   = \frac{-C^{K}_{z'}W_{11}^+ + C^{Q}_{z'}W_{21}^+}{W_{00}}
   \, .
\eeqn

One can get another representation (with $x\to -x$, etc.) for the above
asymmetries by using the following relations among the cross sections that
have opposite in-plane nucleon (target or recoil) polarizations:
\beq
\sigma^R_i\,=\,\sigma^L_{-i}\,, \qquad \sigma^\varphi_i\,=\,
\sigma^{- \varphi}_{-i}\,, \qquad \qquad
 (i = \pm x, ~\pm z, ~\pm x', ~\pm z')
\eeq
and similarly for the primed cross sections (with the polarized final
photon). These relations are due to parity conservation.

Introducing the generic quantities
\beqn
F_1 =& \dis\frac{\sigma^{\pi/4} - \sigma^{-\pi/4}}{2 \bar{\sigma}}
    &= \Sigma_{1x}\zeta_{x}+\Sigma_{1z}\zeta_{z} , \nn
F_2 =& \dis\frac{\sigma^R - \sigma^L}{2 \bar{\sigma}}
    &= \Sigma_{2x}\zeta_{x}+\Sigma_{2z}\zeta_{z} , \nn
F_3 =& \dis\frac{\sigma^\parallel - \sigma^\perp}{2 \bar{\sigma}}
    &= \Sigma_3+\Sigma_{3y}\zeta_{y},
\eeqn
one can write the differential cross section in the following compact form:
\beq\label{sezu}
\frac{d\sigma}{d \Omega}=\frac{d\bar\sigma}{d\Omega}
  \Big\{  1 + \Sigma_y\zeta_y + \vec F \cdot \vec\xi \Big\}  ,
\eeq
which, being supplied with appropriated primes, is valid for any combination
of ingoing or outgoing polarized particles.

Thus, with unpolarized and linearly polarized initial or final photons, one
can access two observables: $d\bar\sigma / d\Omega$ and $\Sigma_3$,
respectively.
When the nucleon (target or recoil) polarization is added, four more
asymmetries appear,
$\Sigma_y$ and $\Sigma_{1x}$, $\Sigma_{1z}$, $\Sigma_{3y}$
(and their primed companions), all
of which vanish below the pion threshold.  When circularly polarized
photons are used with polarized nucleons, two more asymmetries appear,
$\Sigma_{2x}$ and $\Sigma_{2z}$,
both of which survive below the pion threshold.

It is important to emphasize that the formulas like
(\ref{cs-generic}) and (\ref{sezu}), being used for polarized
particle(s) in the final state, give the total yield of the particles
and their polarization matrix density.  To get a partial cross section
of a particle produced with specific final polarization(s), one must
use an average instead of a sum over polarizations. That is equivalent
to inserting the statistical factor $1/g$ into these formulas, where
$g=1,2,2,4$ for the cases (\ref{4reactions}-a,b,c,d), respectively.

\section{Low-energy expansions}
\subsection{Invariant amplitudes and polarizabilities}

A very general method for obtaining low-energy expansions of physical
amplitudes for different reactions consists in introducing invariant
amplitudes free from kinematical singularities and constraints and expanding
them in a power series \cite{abar68,bard68}.
Following this method, we first decompose the invariant amplitudes
$A_i$ into Born and non-Born contributions,
\beq\label{decA}
  A_i(\nu,t)=A_i^{\rm B}(\nu,t) + A_i^{\rm NB}(\nu,t)
      \qquad (i = 1,\ldots 6).
\eeq
The Born contribution is associated with pole diagrams involving a
single nucleon exchanged in $s$- or $u$-channels and $\gamma NN$
vertices taken in the on-shell regime,
\beq
  \Gamma_\mu(p',p)=q\gamma_\mu -
    \frac{\kappa}{4m}[\gamma_\mu,\gamma\cdot(p'-p)].
\eeq
It is completely specified by the mass $m$, the electric charge $eq$, and the
anomalous magnetic moment $e\kappa /2m$ of the nucleon:
\beq
\label{ABorn}
  A_i^{\rm B}(\nu,t) =\frac{m e^2 r_i}{(s - m^2)(u - m^2)}
  = -\frac{e^2 r_i}{4m\omega\omega'},
\eeq
where $e$ is the elementary charge, $e^2 \simeq 4\pi/137$, and
$q=1,0$ for the proton and neutron, respectively.
The pole
residues $r_i$ read
\beqn
  &\dis r_1 =    -2q^2     + r_3\,\frac{t}{4m^2}\,, \quad
        r_2 = 2q\kappa + 2q^2 + r_3\,\frac{t}{4m^2}\,, \nn
  &  r_3 = r_5 = \kappa^2 + 2q\kappa, \quad
     r_4 = \kappa^2,  \quad   r_6 = -2q^2 - r_3.
\eeqn
The Born contribution to $T_{fi}$ possesses all the symmetries of
the total amplitude $T_{fi}$, including gauge invariance, and takes
all singularities of $T_{fi}$ at low energies.

The non-Born parts $A_i^{\rm NB}$ of the invariant amplitudes, which
contain all the structure-dependent information, are regular
functions of $\nu^2$ and $t$ and can be expanded as a power series in
$\nu^2$ and $t$.
Since
\beq
   t= -2\omega \omega'(1 - z), \qquad
   \nu^2 =\omega\omega' + \frac{t^2}{16 m^2}\,,
\eeq
such an expansion can be recast as a power series in the cross-even parameter
$\omega\omega'$:
\beq\label{A-exp}
  A_i^{\rm NB}(\nu,t) = a_{i} + \omega\omega'
     (a_{i,\nu} - 2(1 - z) a_{i,t}) + \cdots .
\eeq
Here the low-energy constants
\beq\label{Ac}
  a_i = A_i^{\rm NB}(0,0), \qquad
  a_{i,\nu} = \Big(\frac{\partial A_i^{\rm NB}}
    {\partial\nu^2}\Big)_{\nu=t=0} \,,
  \qquad  a_{i,t} =
   \Big(\frac{\partial A_i^{\rm NB}}
    {\partial t}\Big)_{\nu=t=0} \,, ~ \ldots
\eeq
parametrize the structure of the nucleon as seen in its two-photon
interactions.

The expansion (\ref{A-exp}) directly leads to a corresponding
low-energy expansion of the total amplitude $T_{fi}$
in the Lab frame.  We first consider the spin-independent
part of Eq.~(\ref{Tlab}).  The spin-independent
part of the Born term $T_{fi}^{\rm B}$ follows from Eq.~(\ref{Tlab}):
\beqn\label{T-B-nospin}
   \frac{N(t)}{8\pi m} T_{fi}^{\rm B,\,nospin}  &=&
    r_0\,\See\Big\{ -q^2 + \frac{\omega\omega'}{4m^2}
    \Big[ \kappa^2 + 2q\kappa - q^2 (1-z)\Big]\,\Big\} \nn && {}
     + r_0\,\Sss \frac{\omega\omega'}{4m^2}
    \Big[ q^2 - z(\kappa + q)^2 \Big].
\eeqn
Here $r_0 = e^2/4\pi m$. The leading term in (\ref{T-B-nospin})
reproduces the Thomson limit,
\beq\label{Thom}
   \frac{1}{8\pi m} T_{fi}^{\rm Thomson} = - r_0 q^2 \,\See.
\eeq

The non-Born contribution to $T_{fi}$ is determined by the structure
constants introduced in the expansion (\ref{A-exp}).
Its spin-independent part starts with a $\omega\omega'$ term,
\beq\label{T-NB-nospin}
  \frac{1}{8\pi m} T_{fi}^{\rm NB,\,nospin} = \omega\omega'
    ( \alpha_E\,\See + \beta_M\,\Sss ) +
    \O(\omega^2\omega^{\prime\,2}).
\eeq
Here the constants $\alpha_E$ and $\beta_M$,
\beq\label{pol}
 4\pi\alpha_E = -a_3 - a_6 - a_1, \quad
 4\pi \beta_M = -a_3 - a_6 + a_1,
\eeq
are identified as the dipole electric and magnetic polarizabilities
of the nucleon, just in accordance with an
$\O(\omega^2)$ effective dipole interaction
\beq
\label{H-eff-aE,bM}
 H_{\rm eff}^{(2),\,\rm nospin} = -\frac12 \, 4\pi
    (\alpha_E \vec E^2 + \beta_M \vec H^2 )
\eeq
of the nucleon with external electric and magnetic fields, leading
to the amplitude (\ref{T-NB-nospin}).  Due to its interference
with the Thomson amplitude (\ref{Thom}), the contribution of the
polarizabilities, Eq.~(\ref{T-NB-nospin}), results in a $\O(\omega^2)$
effect in the differential cross section $d\bar\sigma/d\Omega$ in the
case of the proton ($q = 1$) and in a $\O(\omega^4)$ effect in the case of
the neutron ($q = 0$).

The $\O(\omega^4)$ terms in Eq.~(\ref{T-NB-nospin}) are
also easily read out from Eqs.~(\ref{Tlab}) and (\ref{A-exp}). They are
determined by the dipole polarizabilities and the following
combinations of $\nu$- and $t$-derivative constants $a_{i,\nu}$ and
$a_{i,t}$ of the amplitudes $A_1$, $A_3$ and $A_6$:
\beqn
\label{alpha-nu-t-def}
   4\pi\alpha_\nu &=& -a_{3,\nu}-a_{6,\nu}-a_{1,\nu}-\frac{a_5}{m^2}, \nn
   4\pi \beta_\nu &=& -a_{3,\nu}-a_{6,\nu}+a_{1,\nu}+\frac{a_5}{m^2}, \nn
   4\pi\alpha_t &=& -a_{3,t}-a_{6,t}-a_{1,t}+\frac{a_3}{4m^2}, \nn
   4\pi \beta_t &=& -a_{3,t}-a_{6,t}+a_{1,t}+\frac{a_3}{4m^2}.
\eeqn
These polarizability-like quantities
are constant coefficients of energy- and 
angle-dependent corrections to the dipole interaction (\ref{H-eff-aE,bM})
which enter
to next order in photon energy.
The recoil correction $\sim 1/m^2$
in Eqs.~(\ref{alpha-nu-t-def}) is
explained in Appendix~C.

We now introduce linear combinations of
the parameters
(\ref{alpha-nu-t-def}) which have a more
direct physical meaning.  If we consider the partial-wave structure of the
amplitude $T_{fi}$ (see Appendix~A), we can relate the $t$-derivative
constants in Eq.~(\ref{alpha-nu-t-def}) to the quadrupole
polarizabilites of the nucleon \cite{rade79}:
\beq
  \alpha_{E2}=12\alpha_t, \quad  \beta_{M2}=12\beta_t.
\eeq
A nonrelativistic example is given in Appendix~B.  The quantities
\beq
  \alpha_{E\nu} = \alpha_\nu - 2\alpha_t +  \beta_t, \quad
   \beta_{M\nu} =  \beta_\nu - 2 \beta_t + \alpha_t,
\eeq
which we call ``dispersion polarizabilities", describe the
$\omega$-dependence of the dynamic dipole polarizabilities.  In terms
of the parameters $\alpha_{E2}$, $\beta_{M2}$,
$\alpha_{E\nu}$, and $\beta_{E\nu}$, the corresponding effective
interaction of $\O(\omega^4)$ has the form
\beq
 H_{\rm eff}^{(4),\,\rm nospin} = -\frac12\, 4\pi
    (\alpha_{E\nu} \vec{\dot E}^2 + \beta_{M\nu} \vec{\dot H}^2)
    -\frac1{12}\, 4\pi (\alpha_{E2} E_{ij}^2 + \beta_{M2} H_{ij}^2),
\eeq
where the dots mean a time derivative and
\beq
  E_{ij} = \frac12 (\nabla_i E_j + \nabla_j E_i), \quad
  H_{ij} = \frac12 (\nabla_i H_j + \nabla_j H_i)
\eeq
are the quadrupole strengths of the electric and magnetic fields.

We next consider the spin-dependent part of Eq.~(\ref{Tlab}).
The spin-dependent part of $T_{fi}$ starts with
$\O(\omega)$ terms which come from the Born contribution. To
leading order one has \cite{gell54},
\beqn
\label{TBorn-spin}
   \frac{1}{8\pi m} T_{fi}^{\rm B,\,spin} &=&
   -ir_0\frac{\nu}{2m} \Big( q^2 \Vee + (\kappa+q)^2\Vss \Big)
\nn && \qquad {}   +ir_0 q\frac{\kappa+q}{2m}
    \Big( \omega'\kS\Sse - \omega\Sk\Ses \Big) + \O(\omega^2).
\eeqn
The omitted higher-order terms can
be read out from Eqs.~(\ref{Tlab}) and (\ref{ABorn}).

As shown in Eq.~(\ref{Tlab}), the non-Born
part of $T_{fi}^{\rm spin}$ starts with $\O(\omega^3)$ terms.
They are determined by the four constants $a_2$,
$a_4$, $a_5$, and $a_6$.  Due to their interference with the
spin-dependent $\O(\omega)$ terms of Eq.~(\ref{TBorn-spin}), they
give rise to a $\O(\omega^4)$ correction to the unpolarized cross
section.  In the case of a polarized proton and a circularly polarized
photon, these terms appear at the order $\O(\omega^3)$.

The $\O(\omega^3)$ terms in $T_{fi}^{\rm NB,\,spin}$ were considered in
several papers \cite{lin71,levc85,ragu93}. In the most recent (and best
known) work \cite{ragu93} they are parametrized as
\beqn
\label{gammas-def}
  \frac{1}{8\pi m}\,T_{fi}^{\rm NB,\, spin} &=&
    i\omega^3\gamma_1 \Vee
   +i\omega^3\gamma_2 (\Vkk\See - \Vee\Skk) \nn && \quad{}
   +i\omega^3\gamma_3 (\Ss\Sek - \sS\Ske) \nn && \quad{}
   +i\omega^3\gamma_4 (\Vek\Ske - \Vke\Sek \nn && \qquad\qquad{}
      - 2\Vee\Skk) + \O(\omega^4),
\eeqn
where $\gamma_i$ are structure parameters (often called the
``spin polarizabilities'') which are linear
combinations of the constants $a_i$ (see Appendix A).  In the following
we consider a
linear combination of these parameters
\beqn
  \gamma_{E1} = -\gamma_1 - \gamma_3
        &=& \frac{1}{8\pi m}(a_6 - a_4 + 2a_5 + a_2), \nn
  \gamma_{M1} =  \gamma_4
        &=& \frac{1}{8\pi m}(a_6 - a_4 - 2a_5 - a_2), \nn
  \gamma_{E2} =  \gamma_2 + \gamma_4
        &=& -\frac{1}{8\pi m}(a_4 + a_6 + a_2), \nn
  \gamma_{M2} =  \gamma_3
        &=& -\frac{1}{8\pi m}(a_4 + a_6 - a_2),
\eeqn
which have a more transparent physical meaning,
as explained in Appendices A and B. The
parameters $\gamma_{E1}$ and $\gamma_{M1}$ describe the spin dependence
of the dipole electric and magnetic photon scattering, $E1\to E1$ and
$M1\to M1$, whereas $\gamma_{E2}$ and $\gamma_{M2}$ describe the
dipole-quadrupole amplitudes $M1\to E2$ and $E1\to M2$, respectively.
The amplitude (\ref{gammas-def}) implies an effective
spin-dependent interaction of order $\O(\omega^3)$
\beq
  H_{\rm eff}^{(3),\,\rm spin} =   -\frac12\,4\pi \Big(
     \gamma_{E1} \vec\sigma \cdot \vec E \times \vec{\dot E}
   + \gamma_{M1} \vec\sigma \cdot \vec H \times \vec{\dot H}
   -2 \gamma_{E2} E_{ij}\sigma_i H_j
   +2 \gamma_{M2} H_{ij}\sigma_i E_j \Big).
\eeq

To summarize, we have obtained a low-energy expansion of the amplitudes
for nucleon Compton scattering.  To $\O(\omega^4)$, the
cross section and polarization observables are determined by 10
polarizability parameters:
\begin{itemize}
\item Two dipole polarizabilities:   $\alpha_E$ and $\beta_M$
\item Two dispersion corrections to the dipole polarizabilities:
$\alpha_{E\nu}$ and $\beta_{M\nu}$
\item Two quadrupole polarizabilities:  $\alpha_{E2}$ and $\beta_{M2}$
\item Four spin polarizabilities:  $\gamma_{E1}$,
$\gamma_{M1}$, $\gamma_{E2}$, $\gamma_{M2}$
\end{itemize}
These polarizabilities have a simple physical interpretation in terms
of the interaction of the nucleon with an external electromagnetic
field.  Equivalently, these ten parameters are linear combinations of
the low-energy constants, Eq.~(\ref{Ac}), representing the zero-energy
limit of the six invariant amplitudes $A_i$ plus two combinations of both the
$\nu$- and $t$-devivatives of these amplitudes.

To illustrate the interplay among all these polarizabilities,
we consider the two limiting cases of forward and backward
scattering.

(i) The amplitude for forward scattering,
\beq
\frac{1}{8\pi m}\,\Big[ T_{fi} \Big]_{\theta = 0} =
  \frac{\omega^2}{2\pi} \Big( {-}\See (A_3 + A_6)
    + \frac{\omega}{m}\, i\Vee A_4 \Big),
\eeq
has the following low-energy decomposition:
\beqn
   \frac{1}{8\pi m} \Big[ T_{fi} \Big]_{\theta=0}
    &=& \See\Big( {-}r_0 q^2 + \omega^2(\alpha_E+\beta_M)
   + \omega^4(\alpha_\nu+\beta_\nu) + \O(\omega^6) \Big)
\nn && {} + i\omega\Vee\Big( -\frac{r_0\kappa^2}{2m}
    + \omega^2\gamma + \O(\omega^4) \Big),
\eeqn
where
\beq
\label{gamma-0}
  \gamma = -\gamma_{E1} - \gamma_{M1} - \gamma_{E2} - \gamma_{M2}
          = \frac{a_4}{2\pi m}
\eeq
is the ``forward-angle spin polarizability", and
\beq
   \alpha_\nu + \beta_\nu =
 {\textstyle \alpha_{E\nu}+\beta_{M\nu}+
    \frac1{12}\alpha_{E2} + \frac1{12}\beta_{M2} }
     = -\frac{1}{2\pi}(a_{3,\nu}+a_{6,\nu}).
\eeq

(ii) The amplitude for backward scattering,
\beqn
\frac{1}{8\pi m} \Big[ T_{fi} \Big]_{\theta = \pi}
  &=& -\frac{\omega \omega^{\prime}}{2\pi N(t)}\,
    \Big\{ \See\Big(1-\frac{t}{4m^2}\Big)
       \Big( A_1-\frac{t}{4m^2}A_5 \Big)
 \nn && \qquad   + i\frac{\nu}{m}\Vee
    \Big[ A_2+\Big(1-\frac{t}{4m^2}\Big)A_5 \Big] \,\Big\}
\eeqn
has the following low-energy decomposition:
\beqn
\label{T-pi}
\frac{1}{8\pi m} \Big[ T_{fi} \Big]_{\theta = \pi}  &=&
    N(t)\See \Big\{ {-}r_0q^2 +
       \omega\omega' (\alpha_E-\beta_M)
\nn && \qquad\qquad {}
  + \omega^2\omega^{\prime 2}
   ( \alpha_\nu - \beta_\nu - 4\alpha_t + 4\beta_t )
    + \O(\omega^3\omega^{\prime 3}) \Big\}
\nn && {} + i\sqrt{\omega\omega'}\Vee\Big\{
      \frac{r_0}{2m}(\kappa^2 + 4q\kappa + 2q^2)
     + \omega\omega'\gamma_\pi + \O(\omega^2\omega^{\prime 2}) \Big\},
\eeqn
where 
\beq
\label{gamma-pi}
  \gamma_\pi = -\gamma_{E1} + \gamma_{M1} + \gamma_{E2} - \gamma_{M2}
             = -\frac{a_2+a_5}{2\pi m}
\eeq
is the ``backward-angle spin polarizability", and
\beq
   \alpha_\nu - \beta_\nu - 4\alpha_t + 4\beta_t =
 {\textstyle \alpha_{E\nu}-\beta_{M\nu}-
    \frac1{12}\alpha_{E2} + \frac1{12}\beta_{M2} } =
    \frac{1}{2\pi}\Big(4a_{1,t}-a_{1,\nu} - \frac{a_5}{m^2} \Big).
\eeq

\subsection{Cross sections}

A decomposition analogous to (\ref{decA}) can be applied also to
the invariant functions
$W_{ij}$:
\beq
W_{ij} = W_{ij}^{\rm B} + W_{ij}^{\rm NB} .
\eeq
The Born contribution is given by
\begin{mathletters}
\beqn
\frac{1}{(8\pi m)^2}\,W_{00}^{\rm B}   &=&  \frac{r_0^2}{2}
   \Big\{ q^4 (1 + z^2) + \frac{\omega\omega'}{4m^2}
   \Big[ 4q^3 (q + 2\kappa)(1 - z)^2 + 2q^2 (9 - 10z + z^2)\kappa^2 \nn
 && \qquad\qquad {} + 4q (3 - 2 z - z^2)\kappa^3 +
   (3 - z^2) \kappa^4 \Big] \Big\}, \\
\frac{1}{(8\pi m)^2}\,W_{03}^{\rm B}   &=&  - \frac{r_0^2}{2}
   (1 - z^2)\Big[ q^4 + \frac{\omega\omega'}{4m^2}
   (\kappa^2 + 2q\kappa)^2 \Big], \\
\frac{1}{(8\pi m)^2}\,W_{12}^{\pm,\,\rm B}  &=&
   \frac{r_0^2}{2m}\Big\{ (1 + z) \Big[ q^2 -
   \frac{\omega\omega'}{4m^2} (1 - z) (\kappa^2 + 2q\kappa)\Big]\,
   \Big[\kappa^2 + q(1 - z)(q + \kappa) \Big] \nn
 && \qquad {} \pm \frac{\nu}{2m} (1 - z) (\kappa^2 + 2q\kappa)
   \Big[\kappa^2 + 2q\kappa + q(1 - z) (q + \kappa)\Big]\Big\}, \\
\frac{1}{(8\pi m)^2}\,W_{22}^{\pm,\,\rm B}  &=&
  \frac{r_0^2}{2m}\Big\{ (1 - z) \Big[ q^2 +
  \frac{\omega\omega'}{4m^2} (1 - z) (\kappa^2 + 2q\kappa)\Big]\,
  \Big[\kappa^2 + 2q\kappa + q (1 - z) (q + \kappa)\Big] \nn
 && \qquad {} \mp\frac{\nu}{2m} (1 + z) (\kappa^2 + 2q\kappa)
    \Big[\kappa^2 + q (1 - z) (q + \kappa) \Big]\Big\}.
\eeqn
\end{mathletters}
The non-Born contributions are more conveniently written
by separating the terms of different order in $\omega \omega^{\prime}$.
In particular, for $W_{00}$ and $W_{03}$ we have
\beq
W_{0k}^{\rm NB} = U_{0k}^{(2)}\omega\omega' +
    U_{0k}^{(4)} \omega^2 \omega^{\prime 2} +
 \O(\omega^3\omega^{\prime 3}) \qquad (k = 0,3),
\eeq
where
\begin{mathletters}
\beqn
\frac{1}{(8\pi m)^2}\,U_{00}^{(2)} &=& -r_0 q^2
   \Big[ (1 + z^2)\alpha_E + 2z\beta_M \Big], \\
 \frac{1}{(8\pi m)^2}\,U_{03}^{(2)} &=& r_0 q^2
   (1 - z^2)\alpha_E, \\
\frac{1}{(8\pi m)^2}\,U_{00}^{(4)}  &=&
   \frac12(1 + z^2)(\alpha_E^2 + \beta_M^2)
  + 2z\alpha_E \beta_M +  \frac{r_0}{4m^2}(1-z) \times
\nn && {} \hspace{-3em}
   \Big[ (1 + z) (\kappa^2 + 2q\kappa) (\alpha_E + z\beta_M)
   + 2q^2 (2z\alpha_E + (1+z^2)\beta_M)\Big]
\nn && {} \hspace{-3em}
   - r_0 q^2 \Big[ (1+z^2)\alpha_{E\nu} + 2z\beta_{M\nu}
    +\frac{z^3}{6}\alpha_{E2} + \frac{3z^2-1}{12}\beta_{M2} \Big]
   + \frac{r_0}{2m} P(z), \\
\frac{1}{(8\pi m)^2}\,U_{03}^{(4)}  &=&  (1 -z^2)\Big\{
    -\frac12 (\alpha_E^2 - \beta_M^2)
    - \frac{r_0}{4m^2}(\kappa^2 + 2q\kappa) (\alpha_E + z\beta_M)
\nn && {}
   + r_0 q^2 \Big[ \alpha_{E\nu}
    + \frac{z}{6}\alpha_{E2} - \frac{1}{12}\beta_{M2} \Big]
    + \frac{r_0}{2m} R \Big\}.
\eeqn
\end{mathletters}
Here $P(z)$ and $R$ are polynomials in $z$
of the third and zero order, respectively, which
are determined by the spin polarizabilities as follows:
\beqn
  P(z)  &=&   \Big[ q^2 (1 + 2z - 3z^2)
     - 2q\kappa (1-z)^2 + 2\kappa^2 z \Big] \gamma_{E1}
\nn
        &+&   \Big[ ( q^2 + 2q\kappa) (3 - 2z - z^2)
        + \kappa^2 (3 - z^2) \Big]  \gamma_{M1}
\nn
        &+&   \Big[ -q^2 (1 - 3z^2 + 2z^3)
     - 2q\kappa (1 + z -3z^2 + z^3)
        + \kappa^2 (3z^2-1) \Big] \gamma_{E2}
\nn
        &+&   \Big[ -q^2(1-z)^2
     + 4q\kappa (z - z^2) + 2\kappa^2 z \Big] \gamma_{M2}
\eeqn
and
\beq
   R = q^2 (\gamma_{E1} - \gamma_{M2} )
    - (\kappa+q)^2 (\gamma_{M1} - \gamma_{E2} ).
\eeq

For the invariants $W_{12}$ and $W_{22}$ we can stop the expansion
at $\O(\omega^3)$, since they appear in the squared amplitude
multiplied by terms of $\O(\omega)$. Therefore we have
\beq
\label{W12,truncation}
   W_{k2}^{\pm,\,\rm NB} = U_{k2}^{(2)} \omega\omega' \pm
       U_{k2}^{(3)} \nu\omega\omega' + \O(\omega^2\omega^{\prime 2})
 \qquad (k = 1,2),
\eeq
where
\begin{mathletters}
\beqn
  \frac{1}{(8\pi m)^2}\,U_{12}^{(2)}  &=&  -\frac{r_0}{2m}(1+z)
   \Big[\kappa^2 + q(1-z) (\kappa+q)\Big](\alpha_E + \beta_M)
\nn && \qquad {}
   + r_0 q^2 (1+z) \Big[ \gamma_{E1}+\gamma_{M1}
    + z (\gamma_{E2}+\gamma_{M2}) \Big], \\
  \frac{1}{(8\pi m)^2}\,U_{22}^{(2)}  &=&  -\frac{r_0}{2m}(1-z)
   \Big[\kappa^2 + 2q\kappa + q(1-z) (\kappa+q)\Big](\alpha_E - \beta_M)
\nn && \qquad {}
   - r_0 q^2 (1-z) \Big[ \gamma_{E1}-\gamma_{M1}
    + z (\gamma_{E2}-\gamma_{M2}) \Big], \\
  \frac{1}{(8\pi m)^2}\,U_{12}^{(3)}  &=&  -\frac{r_0}{2m}(1-z)
    \Big\{ \Big[2\kappa^2 + 4q\kappa + q(1-z)(\kappa+q)\Big]
    (\gamma_{E1}-\gamma_{M1})
\nn && \qquad {}
   + \Big[(\kappa^2 + 2q\kappa)(1+z) + q(1-z)(\kappa+q)\Big]
    (\gamma_{E2}-\gamma_{M2}) \Big\}, \\
  \frac{1}{(8\pi m)^2}\,U_{22}^{(3)}  &=&  -\frac{r_0}{2m}(1+z)
    \Big\{ \Big[\kappa^2 + q(1-z)(\kappa+q)\Big]
     \frac{\alpha_E + \beta_M}{m}
\nn && \qquad {}
    + (\kappa+q)\Big[ 2\kappa+q(1-z) \Big] (\gamma_{E1}+\gamma_{M1})
\nn && \qquad {}
    - \Big[ (\kappa+q)^2(1-z) - q\kappa(1+z)\Big]
      (\gamma_{E2}+\gamma_{M2}) \Big\}.
\eeqn
\end{mathletters}

In order to illustrate the interplay between polarizabilities and the
cross sections (in the Lab frame), we consider below the case of
polarized photon beam and polarized target at $\theta=0$,
$\pi/2$, and $\pi$.

(i) Forward scattering:
\beqn
  \Big[ \frac{d\bar\sigma}{d\Omega} \Big]_{\theta=0} &=&
   [r_0 q^2 - \omega^2(\alpha_E+\beta_M)]^2
  + r_0^2\frac{\omega^2}{4m^2}\kappa^4
\nn && {}
  -2r_0 q^2\omega^4 ( {\textstyle \alpha_{E\nu}+\beta_{M\nu}+
    \frac1{12}\alpha_{E2} + \frac1{12}\beta_{M2} } )
  - r_0\frac{\omega^4}{m}\kappa^2 \gamma + \O(\omega^6)
\eeqn
and
\beq
  \Big[ \Sigma_{2z}\frac{d\bar\sigma}{d\Omega} \Big]_{\theta=0} =
   -r_0^2 q^2\frac{\omega}{m}\kappa^2
   +r_0\frac{\omega^3}{m}\kappa^2 (\alpha_E+\beta_M)
   +2r_0 q^2\omega^3\gamma + \O(\omega^5),
\eeq
where $\gamma$ is the forward-angle spin polarizability
(\ref{gamma-0}).

(ii) Backward scattering:
\beqn
  \Big[ \frac{d\bar\sigma}{d\Omega} \Big]_{\theta=\pi} &=&
    \Big( \frac{\omega'}{\omega} \Big)^2  \Big\{  q^2
   \Big( 1+\frac{\omega\omega'}{m^2} \Big)
   \Big[ r_0^2 q^2-2r_0\omega\omega'(\alpha_E-\beta_M) \Big]
  + r_0^2\frac{\omega\omega'}{4m^2}(\kappa^2+4\kappa q+2q^2)^2
\nn && \qquad {}
  + \omega^2\omega^{\prime 2}(\alpha_E-\beta_M)^2
  -2r_0 q^2\omega^2\omega^{\prime 2}
  ( {\textstyle \alpha_{E\nu}-\beta_{M\nu}-
    \frac1{12}\alpha_{E2} + \frac1{12}\beta_{M2} } )
\nn && \qquad {}
  + r_0\frac{\omega^2\omega^{\prime 2}}{m}
   (\kappa^2+4\kappa q+2q^2) \gamma_\pi \Big\} + \O(\omega^6)
\eeqn
and
\beq
  \Big[ \Sigma_{2z}\frac{d\bar\sigma}{d\Omega} \Big]_{\theta=\pi} =
    \nu \Big( \frac{\omega'}{\omega} \Big)^2  \Big\{
    \frac{r_0}{m}(\kappa^2+4\kappa q+2q^2)
    \Big[ r_0 q^2-\omega\omega'(\alpha_E-\beta_M) \Big]
   +2r_0 q^2\omega\omega'\gamma_\pi  \Big\} + \O(\omega^5),
\eeq
where $\gamma_\pi$ is the backward-angle spin polarizability
(\ref{gamma-pi}).

(iii) At $90^\circ$:
\beqn
  \Big[ \frac{d\bar\sigma}{d\Omega} \Big]_{\theta=\pi/2} &=&
    \Big( \frac{\omega'}{\omega} \Big)^2  \Big\{
   \frac{r_0^2}{2} \Big[ q^4 + \frac{\omega\omega'}{4m^2}
    \Big( 3(\kappa+q)^4 - 4\kappa q^3 + q^4 \Big) \Big]
  - r_0 \omega\omega'\alpha_E
    \Big[ q^2 - \frac{\omega\omega'}{4m^2}(\kappa^2+2\kappa q) \Big]
\nn && {}
  + r_0 q^2\frac{\omega^2\omega^{\prime 2}}{2m^2}\beta_M
  + \frac12\omega^2\omega^{\prime 2}(\alpha_E^2 + \beta_M^2)
  - r_0 q^2\omega^2\omega^{\prime 2} (\alpha_{E\nu}-\frac{1}{12}\beta_{M2})
\nn && \hspace{-3em} {}
  + \frac{r_0}{2m}\omega^2\omega^{\prime 2}
  \Big[ {-}(2\kappa q-q^2)\gamma_{E1} + (\kappa+q)^2
    (3\gamma_{M1}-\gamma_{E2}) - q^2\gamma_{M2} \Big] \Big\} + \O(\omega^6),
\eeqn
%
\beqn
  \Big[ \Sigma_3\frac{d\bar\sigma}{d\Omega} \Big]_{\theta=\pi/2} &=&
    \Big( \frac{\omega'}{\omega} \Big)^2  \Big\{
   {-}\frac{r_0^2}{2} \Big[ q^4 + \frac{\omega\omega'}{4m^2}
    (\kappa^2 + 2\kappa q)^2 \Big]
  + r_0 \omega\omega'\alpha_E
    \Big[ q^2 - \frac{\omega\omega'}{4m^2}(\kappa^2+2\kappa q) \Big]
\nn && {}
  - \frac12\omega^2\omega^{\prime 2}(\alpha_E^2 - \beta_M^2)
  + r_0 q^2\omega^2\omega^{\prime 2} (\alpha_{E\nu}-\frac{1}{12}\beta_{M2})
\nn && {}
  + \frac{r_0}{2m}\omega^2\omega^{\prime 2}
  \Big[ q^2 (\gamma_{E1} - \gamma_{M2}) -
 (\kappa+q)^2 (\gamma_{M1} - \gamma_{E2}) \Big]
   \Big\} + \O(\omega^6),
\eeqn
%
\beqn
  \Big[ \Sigma_{2z}\frac{d\bar\sigma}{d\Omega} \Big]_{\theta=\pi/2} &=&
    \Big( \frac{\omega'}{\omega} \Big)^2  r_0\omega \Big\{
    \frac{r_0}{2m} \Big[ \kappa q^3-
       \frac{\omega'}{4m}(\kappa^2+2\kappa q)(\kappa+q)^2 \Big]
\nn && \hspace{-4em} {}
   + \frac{\omega\omega'}{2m}
    \Big[ (\kappa+q)^2\beta_M - \kappa q\alpha_E \Big]
   - \frac{\nu}{4m^2}\omega\omega'
   (\kappa^2+\kappa q+q^2) (\alpha_E + \beta_M)
 - q^2 \omega\omega'\gamma_{E1}
\nn && \hspace{-5em} {}
   + \frac{\nu}{2m}\omega\omega'
    \Big[ \kappa q (\gamma_{E1} - \gamma_{M2})
    - (2\kappa^2+4\kappa q+q^2)\gamma_{M1} 
    + (\kappa+q)^2\gamma_{E2} \Big]
   \Big\} + \O(\omega^5),
\eeqn
%
\beqn
  \Big[ \Sigma_{2x}\frac{d\bar\sigma}{d\Omega} \Big]_{\theta=\pi/2} &=&
    \Big( \frac{\omega'}{\omega} \Big)^2  r_0\omega' \Big\{
   {-}\frac{r_0}{2m} \Big[ q^2(\kappa+q)^2 +
       \frac{\omega}{4m}\kappa^2(\kappa q+2q^2) \Big]
\nn && \hspace{-4em} {}
   + \frac{\omega\omega'}{2m}
    \Big[ (\kappa+q)^2\alpha_E - \kappa q\beta_M \Big]
   + \frac{\nu}{4m^2}\omega\omega'
   (\kappa^2+\kappa q+q^2) (\alpha_E + \beta_M)
    - q^2\omega\omega'\gamma_{M1}
\nn && \hspace{-5em} {}
   - \frac{\nu}{2m}\omega\omega'
    \Big[ \kappa q (\gamma_{M1} - \gamma_{E2})
   - (2\kappa^2+4\kappa q+q^2)\gamma_{E1}
   + (\kappa+q)^2\gamma_{M2} \Big]
   \Big\} + \O(\omega^5).
\eeqn

For the
low-energy expansion of the amplitudes and structure functions, two
further remarks must be considered.
First, as is evident from Eq.~(\ref{Prange}), the $t$-channel
$\pi^0$-exchange contributes only to $T_5$ (and thus only to $A_2$).
Owing to the small pion mass $m_{\pi^0}$, this diagram is a rapidly
varying function of $t$ and therefore it is profitable to keep its
expression unexpanded.  This can be achieved with the following
replacement in the non-Born part of the amplitude $A_2$:
\beq\label{A2NB}
  A_2^{\rm NB}(\nu,t) = A_2^{\pi^0}(t) - A_2^{\pi^0}(0)
     + \bar A_2^{\rm NB}(\nu,t),
\eeq
where
\beq\label{pi0}
   A_2^{\pi^0}(t) = \frac2t T_5^{\pi^0}(t) =
     \frac {g_{\pi NN} F_{\pi^0\gamma\gamma} }{t - m_{\pi^0}^2} \tau_3 ,
\eeq
with $\tau_3=1$ or $-1$ for the proton and neutron, respectively,
and where
\beq
   \Gamma_{\pi^0\to\gamma\gamma} = \frac{m_{\pi^0}^3}{64\pi}
    F^2_{\pi^0\gamma\gamma}
\eeq
is the $\pi^0$ two-photon decay width.  The relative sign of
the $\pi NN$ coupling $g_{\pi NN}$ and the $\pi^0\gamma\gamma$ coupling
$F_{\pi^0\gamma\gamma}$ is negative.  In Eq. (\ref{A2NB}), $\bar
A_2^{\rm NB}$ is a smoother function of $\nu,t$
than $A_2^{\rm NB}$ and is better
approximated by a polynomial in $\omega\omega'$ (which is just the
constant $a_2$ to the order we consider).  In terms of the spin
polarizabilities, this means that the following substitutions have to
be done in all the expansions:
\beqn
  & \gamma_{E1} \to  \gamma_{E1} + X_{2a} f(t), \qquad
  & \gamma_{E2} \to  \gamma_{E2} - X_{2a} f(t), \nn
  & \gamma_{M1} \to  \gamma_{M1} - X_{2a} f(t), \qquad
  & \gamma_{M2} \to  \gamma_{M2} + X_{2a} f(t),
\eeqn
where
\beq
   X_{2a} = \frac{A_2^{\pi^0}(0)} {8\pi m}, \quad
     f(t) = \frac{t}{m_{\pi^0}^2 - t} .
\eeq
Generally, after the separation of the $\pi^0$-exchange contribution, all
the low-energy expansions discussed in this section become valid provided
$\omega\omega'/m_\pi^2$ is a small parameter. The radius of convergence of
the $\omega\omega'$-series is, up to a small correction of
order $\O(m_\pi/m)$, equal to
\beq\label{converge}
    |\omega|  \le  m_\pi
\eeq
and is determined by the pion photo-production threshold where the
amplitudes have a singularity and acquire an imaginary part. Another close
singularity is due to the $t$-channel exchange of two pions, which gives
the same restriction (\ref{converge}).

The second remark is that the low-energy expansions of the structure
functions $W_{ij}$, when used within the radius of convergence, give an
accurate account of the dependence of the cross sections on the dipole
polarizabilities $\alpha_E$, $\beta_M$ and the spin polarizabilities
$\gamma_{E1}$, $\gamma_{M1}$, $\gamma_{E2}$, $\gamma_{M2}$. In the
expressions for $W_{12}^\pm$ and $W_{22}^\pm$, the spin polarizabilities
enter to leading order $\O(\omega\omega')$ as an interference with the
Thomson amplitude (\ref{Thom}).  The subleading terms of order
$\O(\nu\omega\omega')$ appear due to interference with the amplitude
(\ref{TBorn-spin}).  Numerically the latter terms are enhanced by the
anomalous magnetic moment (for example, $\kappa^2 + 4q\kappa + 2q^2 =
12.4$ in Eq.~(\ref{T-pi}) for the proton) and are as important as the
leading terms.  On the other hand, in the expansions for $W_{ij}$, the
leading order contribution of the quadrupole and dispersion
polarizabilities $\alpha_{E2}$, $\beta_{M2}$, $\alpha_{E\nu}$,
$\beta_{M\nu}$ are already of order $\O(\omega^2\omega'^2)$, which is
the highest order included in the expansion.  Therefore the subleading
terms which describe the interference of these polarizabilities with
the amplitude (\ref{TBorn-spin}) are not included, although they may be
comparable in magnitude to the leading terms.  For this reason, it is
generally more accurate to use low-energy expansions for the invariant
amplitudes $A_i(\nu,t)$, including all ten polarizabilities which
appear to the order considered, and then to calculate the structure
functions $W_{ij}$ through Eqs.~(\ref{Wij}).  This more accurate technique
will be used when we discuss the low-energy observables in Section~V.

In principle, all the polarizabilities we have defined can be
determined from experiment.  For example, in the proton case once the dipole
polarizabilities $\alpha_E$ and $\beta_M$ have been determined
(from low-energy experiments that are sensitive only to terms
of order $\O(\omega^2)$), the angular behavior of the coefficients
$W_{12}$ and $W_{22}$ (from the measurements of $\Sigma_{2x}$ and
$\Sigma_{2z}$) enables a determination of all the spin polarizabilities
$\gamma_{E1}$, $\gamma_{M1}$, $\gamma_{E2}$, $\gamma_{M2}$.  Then the
angular dependence of the unpolarized cross section allows one to
determine the remaining polarizabilities $\alpha_{E2}$, $\beta_{M2}$,
$\alpha_{E\nu}$, $\beta_{M\nu}$.
In the next section, we investigate the feasibility of this approach.

\section{Discussion}

In this section we summarize some theoretical predictions for the
polarizabilities, examine the question of convergence of the low-energy
expansion, and discuss the sensitivity of various observables to these
polarizabilities.

\subsection{Theoretical predictions for polarizabilities}

We now consider various theoretical predictions for the
polarizabilities in order to establish the accuracy which should be
achieved in experiments to make them useful for resolving theoretical
ambiguities.  We restrict ourselves to estimates based on chiral
perturbation theory and dispersion relations, since these have proved
to be successful in applications to strong and electromagnetic
interactions of hadrons at low energies.

In its standard form, chiral perturbation theory gives leading-order
(LO) predictions for the polarizabilities entirely in terms of the pion
mass $m_\pi$, the axial coupling constant of the nucleon, $g_A=1.2573
\pm 0.0028$, and the pion decay constant $f_\pi=92.4 \pm 0.3$ MeV.  The
nucleon mass $m$ which describes recoil corrections does not enter into
the LO predictions but appears in the next-to-leading order together
with some additional parameters.  The LO-contributions dominate in the
chiral limit $m_\pi\to 0$ and are expected to provide semi-quantitative
estimates in the real world.  All results quoted here were obtained by
Bernard \ea, as summarized in Ref.\ \cite{bern95}.  Using explicit
formulas from that work, we have
\beqn
\label{leading-order}
    & \alpha_E^{(\rm LO)} = 10 X_1, \qquad
       \beta_M^{(\rm LO)} =    X_1,
\nn
    & \alpha_{E\nu}^{(\rm LO)} = \frac34 X_3, \qquad
       \beta_{M\nu}^{(\rm LO)} = \frac76 X_3,
\nn
    & \alpha_{E2}^{(\rm LO)} =  7 X_3, \qquad
       \beta_{M2}^{(\rm LO)} = -3 X_3,
\nn
    & \gamma_{E1}^{(\rm LO)} = -5X_2 + X_{2a}, \qquad
      \gamma_{M1}^{(\rm LO)} =  -X_2 - X_{2a},
\nn
    & \gamma_{E2}^{(\rm LO)} =   X_2 - X_{2a}, \qquad
      \gamma_{M2}^{(\rm LO)} =   X_2 + X_{2a}.
\eeqn
Here the quantities $X_i$ are proportional to $i$-th powers of the
inverse pion mass:
\beqn
  & \dis
     X_1=\frac{E^2}{24m_\pi} = 1.23 \cdot 10^{-4}\,\mbox{fm}^3,
\qquad
     X_2=\frac{E^2}{12\pi m_\pi^2} = 1.11 \cdot 10^{-4}\,\mbox{fm}^4,
\nn & \dis
     X_{2a}=\frac{E^2\tau_3}{\pi g_A m_{\pi^0}^2}
      = \pm 11.3 \cdot 10^{-4}\,\mbox{fm}^4,
\qquad
     X_3=\frac{E^2}{20m_\pi^3} = 2.96 \cdot 10^{-4}\,\mbox{fm}^5,
\eeqn
where $E=eg_A\sqrt 2/(8\pi f_\pi)$ is the threshold photoproduction
amplitude for charged pions in the chiral limit. Since the values $X_1$,
$X_2$, $X_3$ are determined by pion loops with charged pions, we use
the charged pion mass $m_\pi=139.57$ MeV for obtaining numbers for
these $X$'s.  The terms with $X_{2a}$ describe the $\pi^0$ exchange
(\ref{pi0}) with couplings fixed by the chiral anomaly
(Wess-Zumino-Witten) and by the Goldberger-Treiman relation:
\beq
\label{WZW}
   F_{\pi^0\gamma\gamma} = - \frac{e^2 N_c}{12\pi^2 f_\pi}
  \quad (N_c=3), \qquad  g_{\pi NN} = g_A \frac{m}{f_\pi}.
\eeq
Accordingly, we use the neutral pion mass $m_{\pi^0}=134.97$ MeV to
evaluate $X_{2a}$.%
\footnote
{In principle, the difference between the masses of $\pi^+$ and $\pi^0$
runs beyond the accuracy of leading-order predictions.  Its consistent
treatment has to involve radiative corrections of ${\cal O}(e^2)$.
However, we assume that the use of the experimental masses of the pions
in the present context still makes sense because this is expected to
reduce the size of the counter-term contributions which come from the full
treatment.}

Calculation to the next order(s) involves new parameters
(counter-terms) which are usually estimated via an approximation that
takes into account the nearest resonances and/or loops with strange
particles. Such a procedure may lead to rather different results for
polarizabilities, depending on further details (for example, compare
Ref. \cite{bern95} and Ref. \cite{hols97a}; see also below).  Among the
nucleon resonances, the $\Delta(1232)$ is special because it is
separated from the nucleon by a relatively small energy gap $\Delta =
m_{\Delta(1232)} - m \simeq 2m_\pi$ and has a large coupling to the
$\pi N$ channel.  Therefore this resonance is particularly important
for low-energy phenomena such as the polarizabilities.  In Refs.
\cite{hols97a,hols97b}, the $\Delta(1232)$ was considered as a partner
of the nucleon in the chiral expansion, which was reformulated in terms
of a generic small energy scale $\epsilon = \O(m_\pi,\Delta)$.  In the
following we invoke only one component of such an approach, the
$\Delta(1232)$-pole contribution.  We ignore the contribution of pion
loops with an intermediate $\Delta(1232)$. These terms are relatively
small for the spin polarizabilities, although large for $\alpha_E$ and
$\beta_M$.  We take into account both $M1$ and $E2$ couplings of the
$\Delta$. Although the quadrupole contribution is of higher order in
$\epsilon$, numerically it is not negligible for the spin
polarizability $\gamma_{E2}$.  For the $\Delta$-pole contribution we
have
\beqn
  & \dis
   4\pi\beta_M^{(\Delta)}  =
        \frac{2\mu_{N\Delta}^2}{\Delta}, \quad
   4\pi\beta_{M\nu}^{(\Delta)}  =
        \frac{2\mu_{N\Delta}^2}{\Delta^3}, \quad
   4\pi\alpha_{E2}^{(\Delta)}  =
        \frac{Q_{N\Delta}^2}{2\Delta}, \quad
\nn & \dis
   4\pi\gamma_{M1}^{(\Delta)}  =
        \frac{\mu_{N\Delta}^2}{\Delta^2}, \quad
   4\pi\gamma_{E2}^{(\Delta)}  =
        \frac{-\mu_{N\Delta} Q_{N\Delta}}{2\Delta}
\eeqn
(and nothing for the other polarizabilities, as discussed in detail in
Appendix~D).  Here $\mu_{N\Delta}$ and $Q_{N\Delta}$ are the magnetic
and quadrupole transition moments, respectively.  Depending on how the
$M1$-coupling is extracted from the experimental radiative width of the
$\Delta$ (with relativistic or nonrelativistic phase space), the
$\Delta$-pole contribution to $\beta_M$ was evaluated in
\cite{hols97a,hols97b} as
\beq
\label{betaDelta}
  \beta_M^{(\Delta)} = 12.0~\cite{hols97a}
\quad{\rm or}\quad      7.2~\cite{hols97b}
\eeq
(units are $10^{-4}\,\mbox{fm}^3$).%
\footnote{This difference also illustrates how estimates of
counter-terms may depend on assumptions.}
With the generally accepted phenomenological magnitude of $\mu_{N\Delta}$
(see, for example, \cite{ericson}), $\beta_M^{(\Delta)} \simeq
(12{-}13)$ \cite{petr81,mukh94}. In the Skyrme model \cite{witt83} or
large-$N_c$ QCD \cite{manoh94}, the transition magnetic moment is
related to the isovector magnetic moment of the nucleon as
$\mu_{N\Delta}=\sqrt 2 \mu_N^V=3.33\, (e/2m)$, 
so that $\beta_M^{(\Delta)} = 12.0$.
All these contributions are summarized in Table~I, where the larger
magnitude of Eq.~(\ref{betaDelta}) has been assumed.  With the fixed ratio
$Q_{N\Delta}/\mu_{N\Delta}=-0.25$ fm (see Appendix~D), the use of the
other value would reduce all the $\Delta$-pole contributions by 40\%.

\begin{table}[htbp]  
\caption{
Polarizabilities of the nucleon given by chiral perturbation theory to
leading order in $m_\pi$ (the columns labeled $\pi^0$ and ``loop"
\protect\cite{bern95,babu97}).  Also the $\Delta$-pole contribution is
given with the larger strength of Eq.~(\ref{betaDelta}) (see discussion
in text).  Other predictions are based on dispersion relations
\protect\cite{lvov97,drec97}.  Separately given are the
$\pi^0$-exchange contribution ($A_2^{\rm as}$) and the contribution of
excitations. The set HDT \protect\cite{drec97} uses pion
photoproduction multipoles of Hanstein \ea \protect\cite{hans97}.  The
set SAID is the result of the present work based on the solution SP97K
\protect\cite{arnd96}.  In the columns $\pi^0$ and $A_2^{\rm as}$, the
proton or neutron case corresponds to $\tau_3=1$ or $-1$, respectively.
}

$$
\begin{array}{ l | r | r | r | r | rr | rr }
\hline
   & \multicolumn{2}{c|} {\mbox{CHPT}}
   & \mbox{$\Delta$-pole}
   & \multicolumn{5}{c}  {\mbox{Dispersion Relations}}
\\
   &  \multicolumn{2}{c|} { \mbox{(leading order)}}
   && \multicolumn{1}{c}  { A_2^{\rm as}}
   &  \multicolumn{4}{c}  { {\rm excitations}+A_1^{\rm as} }
\\
   &  \multicolumn{2}{c|}{}
   && \multicolumn{1}{c} {}
   &  \multicolumn{2}{c} {\mbox{proton}}
   &  \multicolumn{2}{c} {\mbox{neutron}}
\\
   & \multicolumn{1}{c}{\pi^0}
   & \mbox{loop} && \multicolumn{1}{c} {}
   & \mbox{HDT}  &  \multicolumn{1}{r} {\mbox{SAID}}
   & \mbox{HDT}  &  \mbox{SAID}
\\ \hline
%
  \alpha_E       &          & 12.3  &       &
    &        &  11.9 ^{a,c}
    &        &  13.3 ^{b,c} \\
  \beta_M       &          &  1.2  & 12.0  &
    &        &   1.9 ^{a,c}
    &        &   1.8 ^{b,c} \\
(10^{-4}\,{\rm fm}^3)  &&&&&&
\\ \hline
 \alpha_{E\nu} &          &  2.2  &       &       &&  -3.8  &&  -2.4  \\
  \beta_{M\nu} &          &  3.5  &  5.3  &       &&   9.1  &&   9.2  \\
 \alpha_{E2}   &          & 20.7  &  0.2  &       &&  27.5  &&  27.2  \\
  \beta_{M2}   &          & -8.9  &       &       && -22.4  && -23.5  \\
 (10^{-4}\,{\rm fm}^5) &&&&&& \\
\hline
\gamma_{E1}&  11.3 \tau_3 & -5.5 &
           &  11.2 \tau_3 & -4.5 & -3.4 & -5.5 & -5.6\\
\gamma_{M1}& -11.3 \tau_3 & -1.1 &  4.0
           & -11.2 \tau_3 &  3.4 &  2.7 &  3.4 &  3.8\\
\gamma_{E2}& -11.3 \tau_3 &  1.1 & 0.75
           & -11.2 \tau_3 &  2.3 &  1.9 &  2.6 &  2.9\\
\gamma_{M2}&  11.3 \tau_3 &  1.1 &
           &  11.2 \tau_3 & -0.6 &  0.3 & -0.6 & -0.7\\
  \rule{0ex}{3ex}
\gamma     &              &  4.4 & -4.8
           &              & -0.6 & -1.5 & -0.2 & -0.4\\
\gamma_\pi & -45.3 \tau_3 &  4.4 &  4.8
           & -45.0 \tau_3 & 10.8 &  7.8 ^d & 12.1 & 13.0\\
 (10^{-4}\,{\rm fm}^4) &&&&&& \\
\hline
\multicolumn{9}{l}   {\mbox{\footnotesize
   $^a$   Experimental values for the proton are
$\alpha_E$=12.1$\pm$1, $\beta_M$=2.1$\mp$1 \cite{macg95}.}} \\
\multicolumn{9}{l}   {\mbox{\footnotesize
   $^b$   Experimental values for the neutron are
$\alpha_E$=12.6$\pm$2.5 \cite{schm91} and 0$\pm$5 \cite{koes95}. }} \\
\multicolumn{9}{l}   {\mbox{\footnotesize
\quad 
See also the criticism of the former result in Ref.~\cite{enik97}. }} \\
\multicolumn{9}{l}   {\mbox{\footnotesize
   $^c$
$\alpha_E-\beta_M$=10.0 and 11.5 is used as input for the proton and neutron,
respectively.}} \\
\multicolumn{9}{l}   {\mbox{\footnotesize
\quad This guarantees that $A_1^{\rm{as}}(0)$ is the same for
the proton and neutron.}} \\
\multicolumn{9}{l}   {\mbox{\footnotesize
   $^d$ Experimental value reported for the non-$\pi^0$ part of
$\gamma_\pi$ for the proton is 17.9$\pm$3.4 \cite{legs97a}.}}
\end{array}
$$
\end{table}

Another way to calculate the polarizabilities is provided by dispersion
relations for the amplitudes $A_i(\nu,t)$. In the following we give
results of the fixed-$t$ dispersion relations which have the form
\cite{lvov97}
\beq
\label{DR}
  \Re A_i(\nu,t)=A_i^{\rm B}(\nu,t)
   + \frac{2}{\pi} P \int_{\nu_{\rm thr}}^{\nu_{\rm max}}
   \Im A_i(\nu',t)\frac{\nu'd\nu'}{{\nu'}^2-\nu^2}
     + A_i^{\rm as}(t).
\eeq
Here the dispersion integral is taken from the pion photoproduction
threshold to some large energy (actually, it is 1.5 GeV).  The
remaining so-called asymptotic contribution is $t$-dependent but only
weakly energy dependent.  From unitarity, the imaginary part of the
amplitudes $A_i$ and the integral in Eq.  (\ref{DR}) can be calculated
by using experimental information on single- and double-pion
photoproduction which is available at energies below $\nu_{\rm max}$.
In the present context we use the latest version of the single-pion
photoproduction amplitudes provided by the partial wave analysis of the
VPI group \cite{arnd96} (the code SAID, solution SP97K). Also,
double-pion photoproduction is taken into account in the framework of a
model which includes both resonance mechanisms and nonresonant production
of the $\pi\Delta$ and $\rho N$ states.  Details of this procedure are
given in Ref. \cite{lvov97}.

The quantities $A_i^{\rm as}(t)$ take into account contributions of
high energies into the dispersion relations. According to Regge
phenomenology, only the amplitudes $A_1$ and $A_2$ can have a
nonvanishing part at high energies and thus have a large asymptotic
contribution. For the other amplitudes ($A_{3,4,5,6}$), the dispersion
integrals provide a very good estimate of the corresponding non-Born
parts. Thus, we can reliably find through Eq. (\ref{DR}) the constants
$a_{3,4,5,6}$ and $a_{3,t}$, $a_{6,t}$, $a_{3,\nu}$, $a_{6,\nu}$.
Moreover, the constant $a_{1,\nu}$ can be found as well, since it is
not affected by the asymptotic contribution.  Uncertainties appear only
for the constants $a_1$, $a_2$, and $a_{1,t}$, which get a sizable
contribution from the asymptotic part of the amplitude.

Following the arguments of Ref. \cite{lvov97}, we relate the asymptotic
contributions $A_{1,2}^{\rm as}(t)$ to the $t$-channel exchange of the
$\sigma$ and $\pi^0$ mesons, respectively.  The couplings of the
$\pi^0$ are well known, and we include this contribution by virtue of
Eq.\ (\ref{pi0}), in which we use experimental values for couplings
{\it and} introduce a monopole form factor with $\Lambda_\pi = 700$
MeV.  The experimental magnitude of the product $g_{\pi NN}
F_{\pi^0\gamma\gamma} = -0.33\,{\rm GeV}^{-1}$ is $\simeq 3\%$ higher
than that given by the anomaly equation (\ref{WZW}). However, the form
factor reduces the $\pi^0$ strength by $(m_\pi/\Lambda_\pi)^2\simeq
4\%$ and makes the resulting contribution at $t=0$ very close to that
given by the anomaly (see Table~I).  Nevertheless, beyond the
$\pi^0$, the amplitude $A_2$ can have contributions from other heavier
exchanges and thus have an additional piece which weakly depends on
$t$.  \footnote{
A recent analysis \cite{legs97a} of unpolarized Compton scattering data 
suggests the existence of such an additional contribution, since
the spin polarizability $\gamma_\pi$ found there deviates considerably 
from that predicted by
the dispersion integral (\ref{DR}) for the amplitude $A_2$ and by the
$\pi^0$ asymptotic contribution (see Table~I).  
Even after allowing for an
uncertainty of $\sim \pm 2$ in the integral contribution given in
Table I, as suggested by a sizable integrand at
energies above 1 GeV where calculations of $\Im A_i$ are very model
dependent,
such a strong deviation is difficult to understand theoretically and needs
to be confirmed by additional experimental measurements.}
Therefore the fixed-$t$ dispersion relation for $A_2(\nu,t)$ should not
be considered as a reliable source of information on $a_2$.

The amplitude $A_1(\nu,t)$ gets a large asymptotic contribution from
$\sigma$ exchange, which contributes to the constant $a_1$ and
therefore to the difference of the dipole polarizabilities $\alpha_E -
\beta_M$.  Since this difference is experimentally known for the proton
to be $\alpha_E - \beta_M = (10 \pm 2)\cdot 10^{-4}\,{\rm fm}^3$
\cite{macg95}, as determined by low-energy Compton scattering data, the
constant $a_1$ is known too.  However, the quantity $a_{1,t}$ is not
fixed by these data and cannot be unambiguously predicted from the
dispersion relations (\ref{DR}).  Using data on $\gamma p$ scattering
at higher energies (above 400 MeV), the $t$-dependence of the
asymptotic contribution $A_1^{\rm as}(t)$ at $-t \sim 0.5\,{\rm GeV}^2$
was estimated in \cite{lvov97}. It corresponds to a monopole form
factor with the cut-off parameter $M_\sigma \simeq 600$ MeV (``mass" of the
$\sigma$).  However, the $t$-dependence of $A_1^{\rm as}(t)$ at smaller
$t$ might be somewhat steeper as suggested from independent estimates.
Therefore, the fixed-$t$ dispersion relation for $A_1(\nu,t)$ should
not be considered as a reliable source of information on $a_{1,t}$.

With these precautions, we put into Table~I the results of saturating
the dispersion integrals by known photoproduction amplitudes and by the
asymptotic contributions discussed above.  We assume that $A_1^{\rm as}$
is the same for the proton and neutron. Depending on which photoproduction
input is used, the integrals are slightly different. We give both our
results, which use the SAID solution SP97K as an input and a model for
double-pion photoproduction, and the results by Drechsel \ea\
\cite{drec97}, which use photoproduction amplitudes by Hanstein \ea\
(HDT) \cite{hans97} (and ignore the double-pion channel).  The main
difference between these two evaluations comes from near-threshold
energies where the SAID and HDT multipoles $E_{0+}$ are rather
different (see Ref. \cite{drec97} for details).  For some other
(generally similar) dispersion estimates, see Refs.
\cite{levc85,sand94,babu98}. Available experimental data are also given in
the footnotes to
Table~I, including measurements of $\alpha_E$ and $\beta_M$ for
the proton (see Ref. \cite{macg95} and references therein),
measurements of $\alpha_E$ for the neutron \cite{schm91,koes95} (see
also \cite{enik97} for critical discussion), and a recent estimate
of $\gamma_\pi$ for the proton \cite{legs97a}.
It is apparent from Table~I that it is necessary to measure the spin
polarizabilities to an accuracy $\sim 1\cdot 10^{-4}\,{\rm fm}^4$ in order
to obtain constraints valuable for the theory.

Similarly to Ref.\ \cite{drec97}, we can isolate quantities which can
be reliably  predicted by the dispersion relations (\ref{DR})
(up to small uncertainties coming from the photoproduction
amplitudes), namely those
which do not depend on $a_1$, $a_2$, and $a_{1,t}$
(see Table~II).  In the absence of precise data on double-polarized Compton
scattering at low energies, these predictions can be used to diminish
the number of unknown parameters during fits and to help in
determination of unconstrained combinations. These latter combinations are
$\alpha_E-\beta_M$ (which has already been measured), $\gamma_\pi$, and
$\alpha_{E2} - \beta_{M2}$.  Therefore, the most interesting and
informative experiments on Compton scattering are those which are
sensitive to the parameters $\gamma_\pi$ and $\alpha_{E2}-\beta_{M2}$,
which mainly contribute to the backward scattering amplitude.  Below we
discuss which observables are the best suitable to this aim.  
The above remarks apply principally to the proton.
The case
of the neutron can considered in a similar manner.
However, since a free neutron target is not available,
neutron polarizabilities are studied
through elastic or inelastic {\it nuclear} Compton scattering, thereby
introducing additional uncertainites due to 
Fermi motion, meson exchange currents, final-state
interactions, etc. (see, e.g., Refs.\
\cite{rose90,levc94}).

\begin{table}[htbp]  
\caption{
Combinations of polarizabilities of the nucleon which do not depend on
the asymptotic contributions $A_{1,2}^{\rm as}(t)$.  Notation is the
same as in Table~I.}

$$
\begin{array}{ l | r@{\qquad} | r@{\quad} | r r }
\hline \rule{0ex}{2.5ex}
   & \multicolumn{1}{c|}{\mbox{CHPT~(LO)}}
   & \multicolumn{1}{c|}{\mbox{$\Delta$-pole}}
   & \multicolumn{2}{c }{\mbox{DR~(SAID)}}
\\ &&
   & {\mbox{proton}}
   & {\mbox{neutron}}
\\ \hline
  \alpha_E+\beta_M \mbox{ $^a$} & 13.6 & 12.0 & 13.8 & 15.1 \\
  (10^{-4}\,{\rm fm}^3)  &&&
\\ \hline
  \alpha_{E\nu}+ \beta_{M\nu}        &  5.7 & 5.3  &  5.3 &  6.8 \\
  \alpha_{E2}  + \beta_{M2}          & 11.8 & 0.2  &  5.1 &  3.7 \\
  \alpha_{E\nu}+\frac14\alpha_{E2}   &  7.4 & 0.05 &  3.1 &  4.4 \\
  \rule{0ex}{2.5ex}
   \beta_{M\nu}+\frac14 \beta_{M2}   &  1.2 & 5.3  &  3.5 &  3.4 \\
  (10^{-4}\,{\rm fm}^5)  &&&
\\ \hline
  \gamma_{E1}+\gamma_{M1}            & -6.7 & 4.0  & -0.7 & -1.8 \\
  \gamma_{E2}+\gamma_{M2}            &  2.2 & 0.75 &  2.2 &  2.2 \\
  \gamma_{E1}+\gamma_{E2}            & -4.4 & 0.75 & -1.5 & -2.6 \\
  \gamma_{M1}+\gamma_{M2}            &    0 & 4.0  &  3.0 &  3.0 \\
  (10^{-4}\,{\rm fm}^4)  &&&
\\ \hline
\multicolumn{5}{l} {\mbox{\footnotesize
   $^a$ Experimental value for the proton is
$\alpha_E+\beta_M$ = 15.2$\pm$2.6 \cite{macg95}. }}  \\
\multicolumn{5}{l}   {\mbox{\footnotesize
\quad A recent evaluation of this quantity in terms of the Baldin }} \\
\multicolumn{5}{l}   {\mbox{\footnotesize
\quad sum rule yields 13.69 $\pm$ 0.14 (stat) for the proton and }} \\
\multicolumn{5}{l}   {\mbox{\footnotesize
\quad 14.40 $\pm$ 0.66 (stat) for the neutron \cite{babu98}.}}
\end{array}
$$
\end{table}

\subsection{Convergence of the low-energy expansion}

\begin{figure}[hbtp]
\begin{center}
\epsfig{angle=0,width=6in,file=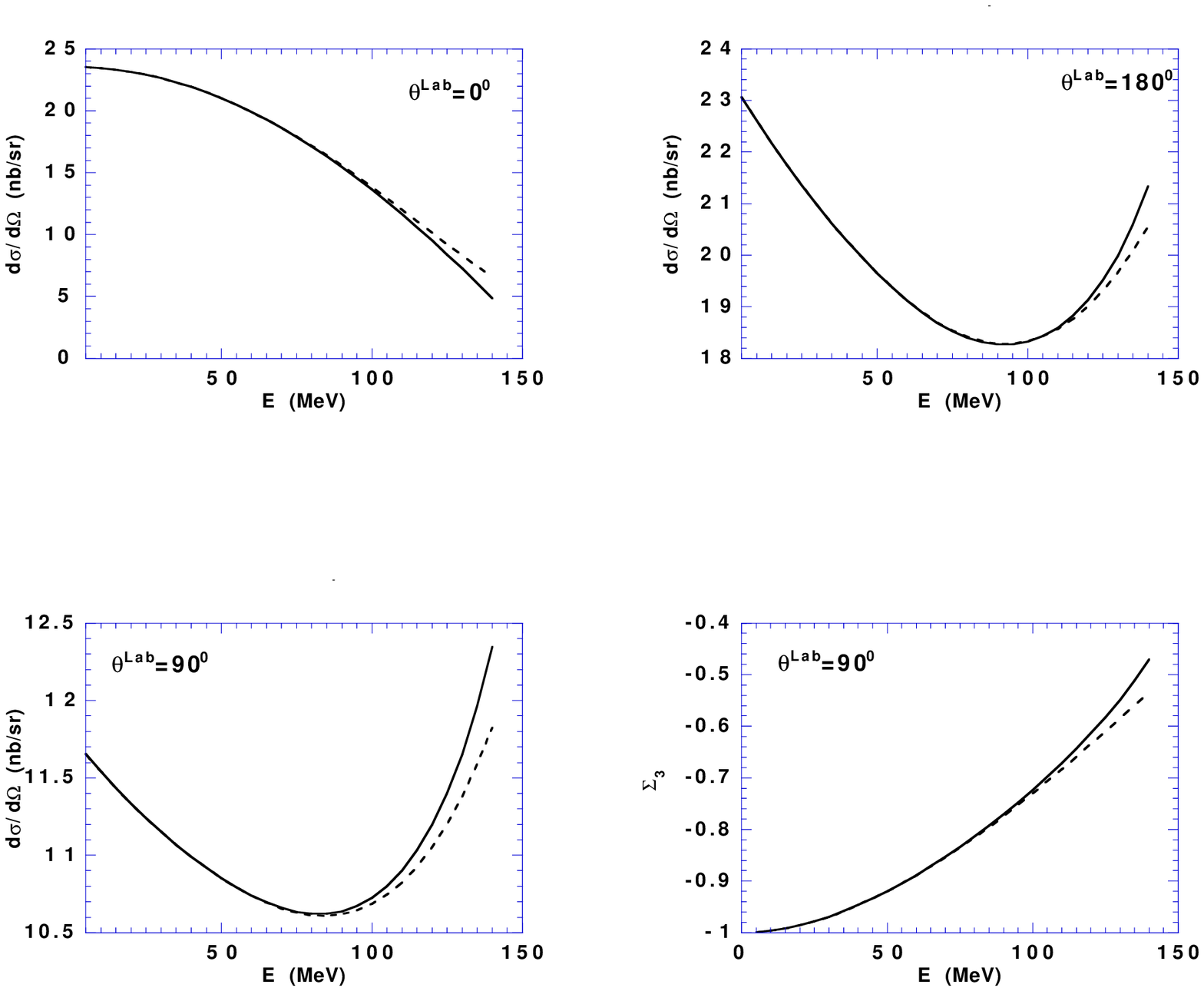}
\vspace{0.25in}
\caption{
Calculated cross sections and asymmetries for the proton.
A comparison is given between a full dispersion calculation
(full curves) and the low-energy expansion (dashed curves)
in which the amplitudes are expanded to $\O(\omega^2\omega'^2)$.
The upper panels are the unpolarized cross section at 
0$^\circ$ (left) and 180$^\circ$ (right).  The lower two panels
are the unpolarized cross section (left) and $\Sigma_3$ (right)
at 90$^\circ$.}
\label{fig:s-p}
\end{center}
\end{figure}

\begin{figure}[hbtp]
\begin{center}
\epsfig{angle=0,width=6in,file=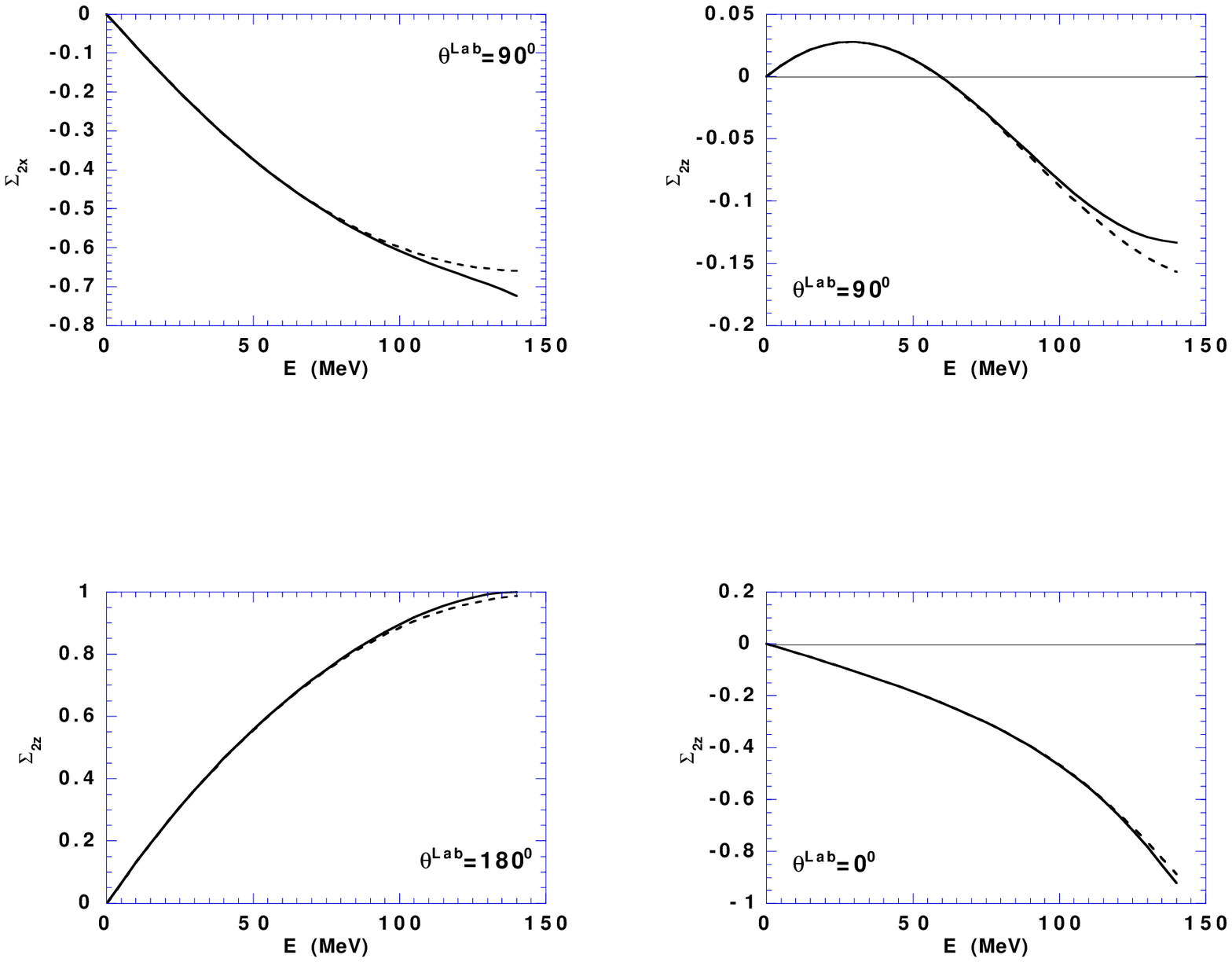}
\vspace{0.25in}
\caption{
Calculated asymmetries for the proton.
A comparison between a full dispersion calculation
(full curves) and the low-energy expansion (dashed curves)
in which the amplitudes are expanded to $\O(\omega^2\omega'^2)$.
The upper panels are $\Sigma_{2x}$ (left) and $\Sigma_{2z}$ (right) at 
90$^\circ$.  The lower two panels are
$\Sigma_{2z}$ at 180$^\circ$ (left) and 0$^\circ$ (right).
}
\label{fig:s2zx-p}
\end{center}
\end{figure}

\begin{figure}[hbtp]
\begin{center}
\epsfig{angle=0,width=6in,file=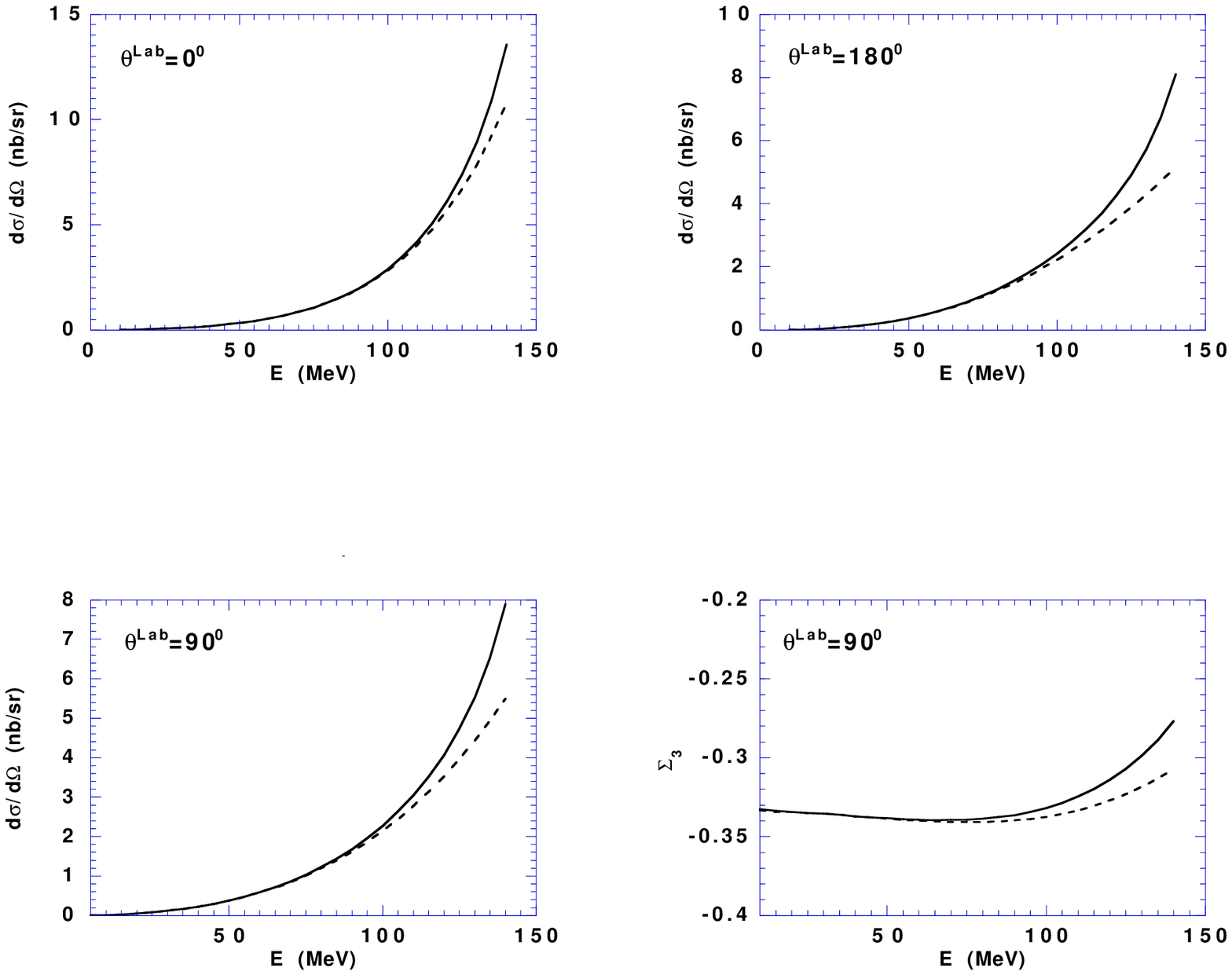}
\vspace{0.25in}
\caption{
Calculated cross sections and asymmetries for the neutron.
A comparison is given between a full dispersion calculation
(full curves) and the low-energy expansion (dashed curves)
in which the amplitudes are expanded to $\O(\omega^2\omega'^2)$.
The upper panels are the unpolarized cross section at 
0$^\circ$ (left) and 180$^\circ$ (right).  The lower two panels
are the unpolarized cross section (left) and $\Sigma_3$ (right)
at 90$^\circ$.}
\label{fig:s-n}
\end{center}
\end{figure}

\begin{figure}[hbtp]
\begin{center}
\epsfig{angle=0,width=6in,file=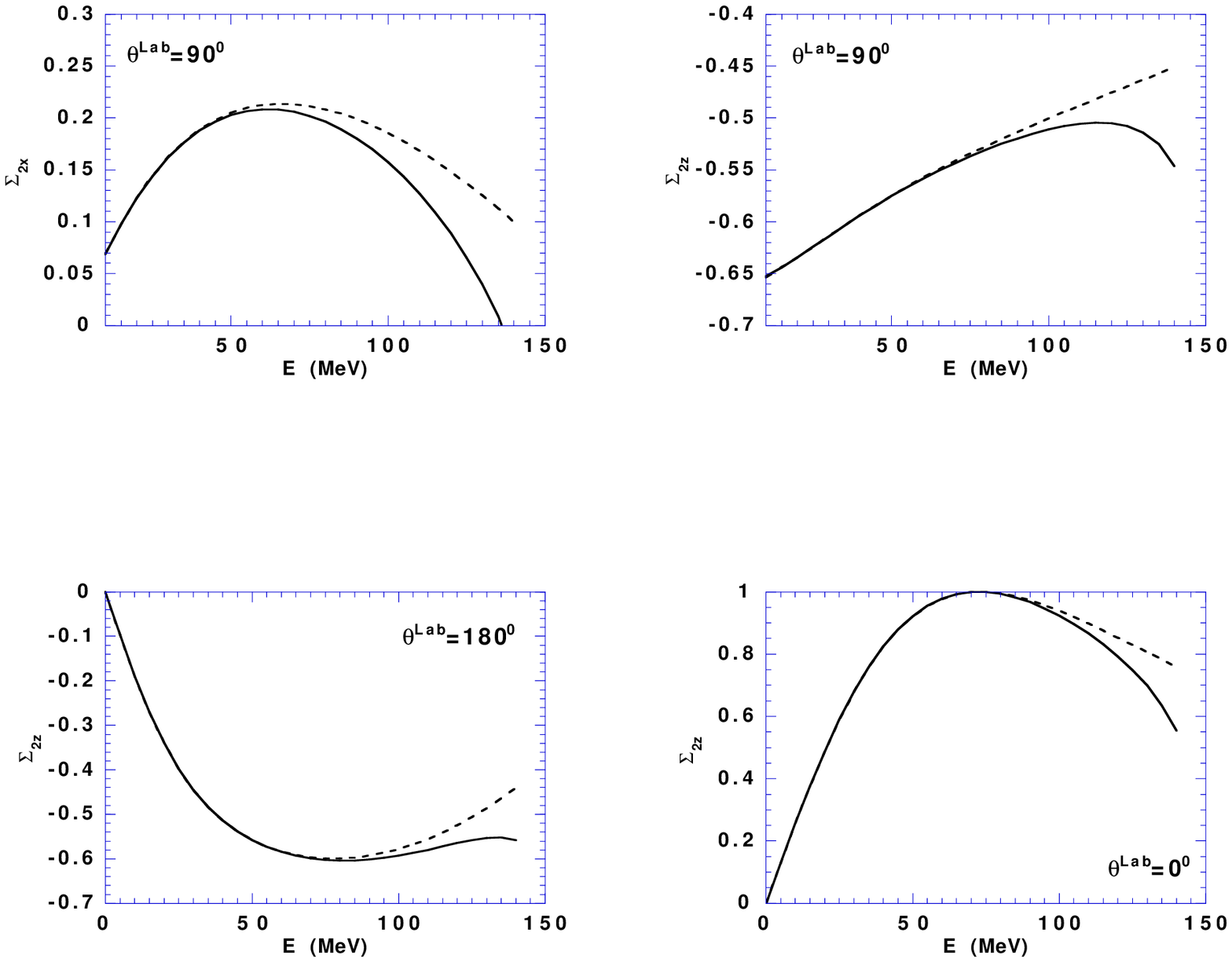}
\vspace{0.25in}
\caption{
Calculated asymmetries for the neutron.
A comparison between a full dispersion calculation
(full curves) and the low-energy expansion (dashed curves)
in which the amplitudes are expanded to $\O(\omega^2\omega'^2)$.
The upper panels are $\Sigma_{2x}$ (left) and $\Sigma_{2z}$ (right) at 
90$^\circ$.  The lower two panels are
$\Sigma_{2z}$ at 180$^\circ$ (left) and 0$^\circ$ (right).
}
\label{fig:s2zx-n}
\end{center}
\end{figure}

In order to investigate the convergence of the low-energy expansion,
we use the same fixed-$t$ dispersion relations described in the preceeding
section to make an exact prediction for the invariant amplitudes
at any energy.  From these the
$W_{ij}$ can be calculated exactly and predictions for the various
observables can be made.  These can be compared with predictions
calculated by truncating the $T_{fi}$ to $\O(\omega^2\omega'^2)$,
as described in Section IV.  Such a comparison is shown in
Figs.~\ref{fig:s-p} and \ref{fig:s2zx-p} for the proton and 
Figs.~\ref{fig:s-n} and \ref{fig:s2zx-n} for the neutron.
For the proton, the expansion works very well for energies up to the pion
threshold.  For the neutron, it appears to work less well due to the fact
that the Born contributions are considerably smaller than for the proton.
The full calculation deviates rapidly from the expansion once the pion
threshold is crossed.

\subsection{Sensitivity of observables to polarizabilities}

\begin{figure}[hbtp]
\begin{center}
\epsfig{angle=0,width=6in,file=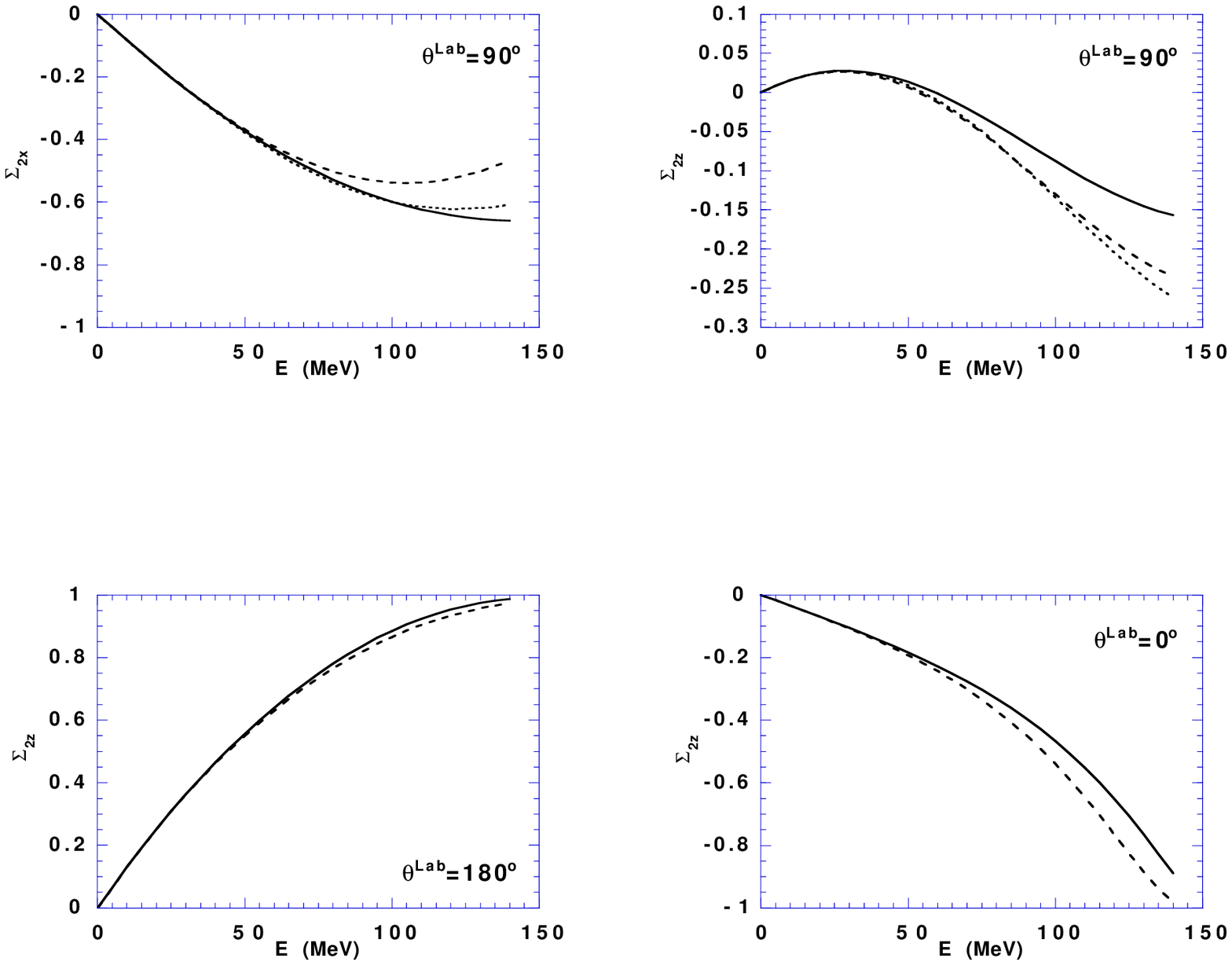}
\vspace{0.25in}
\caption{
Plots for the proton indicating the sensitivity of the double polarization
observables to the spin polarizabilities as a function of energy.
The solid curves use the SAID dispersion values for
all of the spin polarizabilities (Table I).
The upper panels show $\Sigma_{2x}$ (left) and $\Sigma_{2z}$ (right)
at 90$^\circ$.  The long and short dashed curves
increase by $4 \cdot 10^{-4}$ fm$^4$ the values of $\gamma_{E1}$
and $\gamma_{M1}$, respectively.   The lower panels show 
$\Sigma_{2z}$ at 180$^\circ$ (left) and 0$^\circ$ (right).  The dashed
curves decrease by 4 $\cdot$ 10$^{-4}$ fm$^4$ the values of
$\gamma$ (right) or $\gamma_\pi$ (left).
}
\label{fig:sensitivity-p}
\end{center}
\end{figure}

\begin{figure}[hbtp]
\begin{center}
\epsfig{angle=0,width=6in,file=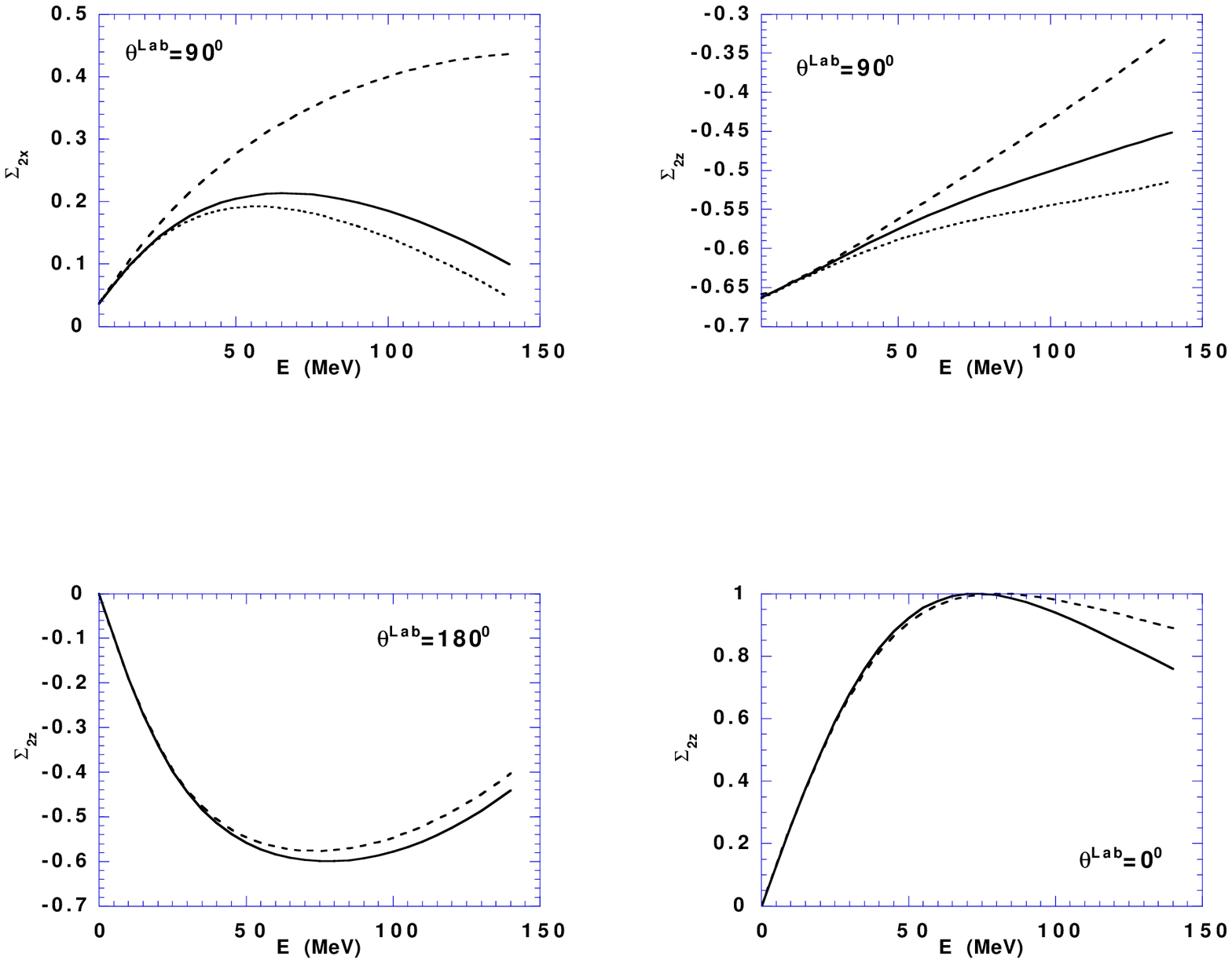}
\vspace{0.25in}
\caption{
Plots for the neutron indicating the sensitivity of the double polarization
observables to the spin polarizabilities as a function of energy.
The solid curves use the SAID dispersion values for
all of the spin polarizabilities (Table I).
The upper panels show $\Sigma_{2x}$ (left) and $\Sigma_{2z}$ (right)
at 90$^\circ$.  The long and short dashed curves
increase by 4 $\cdot$ 10$^{-4}$ fm$^4$ the values of $\gamma_{E1}$
and $\gamma_{M1}$, respectively.   The lower panels show 
$\Sigma_{2z}$ at 180$^\circ$ (left) and 0$^\circ$ (right).  The dashed
curves decrease by 4 $\cdot$ 10$^{-4}$ fm$^4$ the values of
$\gamma$ (right) or $\gamma_\pi$ (left).
}
\label{fig:sensitivity-n}
\end{center}
\end{figure}

We now address the question of the measurement of the polarizabilities
we have defined, with the principal focus being on the spin
polarizabilities.  To the order of our expansion, the double
polarization observables are not sensitive to the quadrupole
and dispersion polarizabilities, whereas the unpolarized
cross section and $\Sigma_3$ are sensitive to all of the
polarizabilities.  Therefore it seems reasonable to use
double polarization measurements to constrain the $\gamma_i$, then
unpolarized measurements to measure the remaining polarizabilities.
At extreme forward and backward angles, the observable $\Sigma_{2z}$
is sensitive to the $\gamma_i$ in the combinations that give
$\gamma$ (Eq.~\ref{gamma-0}) and $\gamma_\pi$ (Eq.~\ref{gamma-pi}),
respectively.  Two other linear combinations can be obtained from
measurements of $\Sigma_{2x}$ and $\Sigma_{2z}$ at 90$^\circ$.
Indeed, the formulas given at the end of Section IV suggest that
these measurements are primarily sensitive to $\gamma_{M1}$
and $\gamma_{E1}$, respectively, at least to the lowest order.
In Figs.~\ref{fig:sensitivity-p} and \ref{fig:sensitivity-n} we 
show these observables as a function
of energy, as calculated using our low-energy expansion with
polarizabilities fixed by the SAID dispersion relation values given
in Table I.   In order to evaluate the sensitivity of the observable
to the polarizability, we have adjusted various quantities by
$4 \cdot 10^{-4}$ fm$^4$ relative to the value in Table I.
This amount is comparable to the pion loop contribution to each of
the polarizabilities but is somewhat larger than the typical 
discrepancy among the competing theories.

We conclude that
in the energy regime below pion threshold where the low-energy
expansion is valid, 
it will be very difficult to measure the spin polarizabilities
to an accuracy that can discriminate among the theories, at least at the
extreme forward and backward angles.  In
particular, the backward spin asymmetry $\Sigma_{2z}$ is almost
completely insensitive to theoretically motivated changes to
$\gamma_\pi$, whereas the forward spin asymmetry is only moderately
sensitive to changes in $\gamma$.  Somewhat more sensitive are the asymmetries
at 90$^\circ$, which might provide some useful constraints on
$\gamma_{E1}$ and $\gamma_{M1}$.  Most promising is $\Sigma_{2x}$ for the
neutron, which is remarkably sensitive to changes in $\gamma_{M1}$.

\begin{figure}[hbtp]
\begin{center}
\epsfig{angle=0,width=6in,file=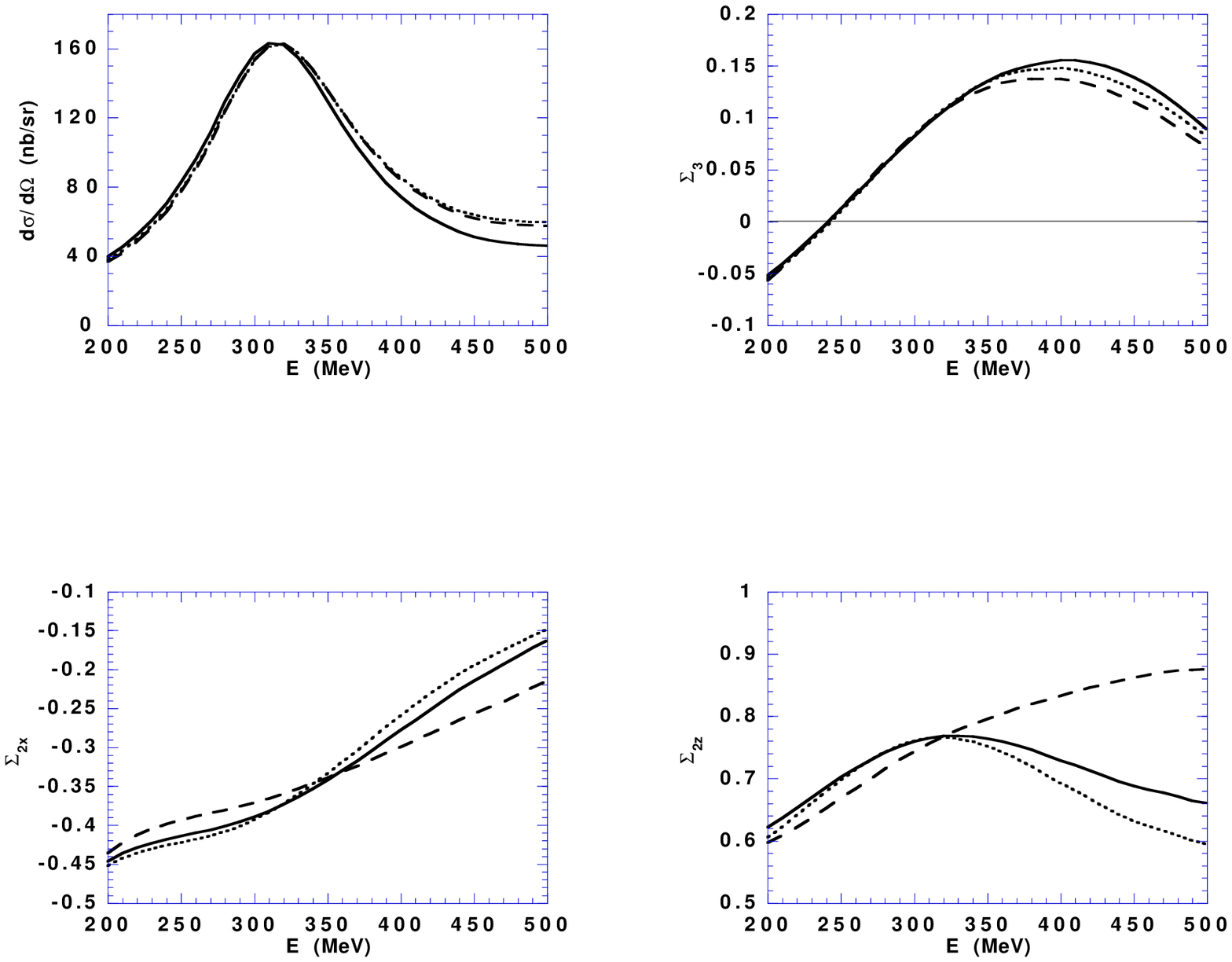}
\vspace{0.25in}
\caption{
Plots for the proton indicating the sensitivity of
$\Sigma_{2x}$ (bottom left),
$\Sigma_{2z}$ (bottom right), $\Sigma_3$ (top right),
and the unpolarized cross section (top left)
to changes
in the the backward spin polarizability $\gamma_\pi$ and the
parameter $a_{1,t}$ at
$\theta^{\rm Lab}=135^\circ$ as a function of energy.
The solid curve uses $a_{1,t}$ corresponding to $M_\sigma$=500 MeV and
$\gamma_{\pi}$=$-37\cdot 10^{-4}~\mbox{fm}^4$; 
the long dashes correspond to 500 MeV and $-41$;
and the short dashes correspond to 700 MeV and $-37$.
}
\label{fig:sigmadelta-p}
\end{center}
\end{figure}

\begin{figure}[hbtp]
\begin{center}
\epsfig{angle=0,width=6in,file=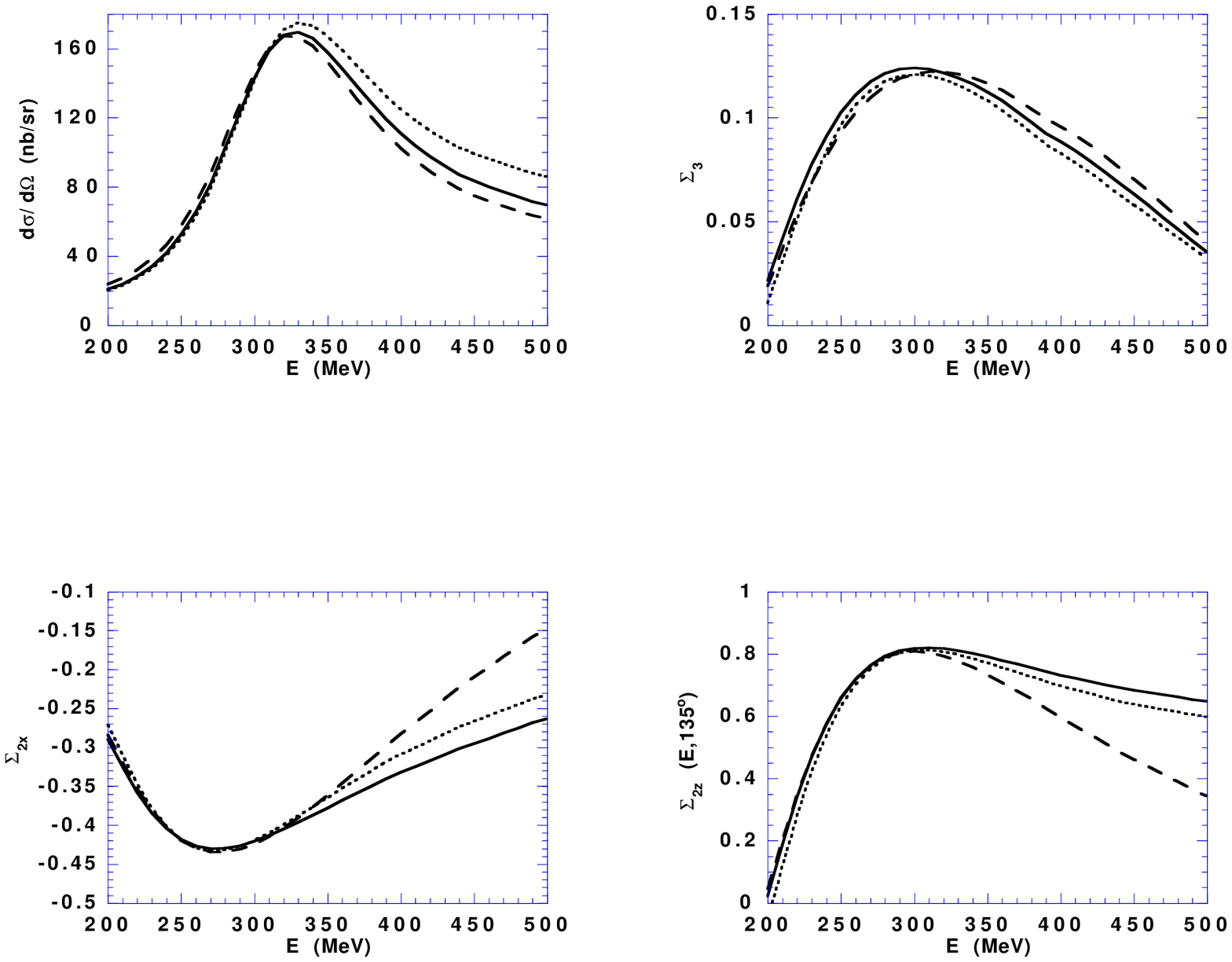}
\vspace{0.25in}
\caption{
Plots for the neutron indicating the sensitivity of
$\Sigma_{2x}$ (bottom left),
$\Sigma_{2z}$ (bottom right), $\Sigma_3$ (top right),
and the unpolarized cross section (top left)
to changes
in the the backward spin polarizability $\gamma_\pi$ and the
parameter $a_{1,t}$ at
$\theta^{\rm Lab}=135^\circ$ as a function of energy.
The solid curve uses $a_{1,t}$ corresponding to $M_\sigma$=500 MeV and
$\gamma_{\pi}$=$58\cdot 10^{-4}~\mbox{fm}^4$; 
the long dashes correspond to 500 MeV and 62;
and the short dashes correspond to 700 MeV and 58.
}
\label{fig:sigmadelta-n}
\end{center}
\end{figure}

At higher energy, the spin asymmetries are more sensitive to
the spin polarizabilities, but of course the low-energy expansion
is no longer valid.  Dispersion theory
provides a convenient formalism for
interpreting Compton scattering data beyond the low-energy approximation,
but only for those polarizabilties not already constrained by the
same dispersion relations.  As discussed above, $\gamma_\pi$
and $\alpha_{E2}-\beta_{M2}$ are not well constrained by the dispersion
relations due to potentially unknown asymptotic
contributions to $a_2$ and $a_{1,t}$, respectively.  We therefore
investigate whether Compton scattering in the so-called dip region between
the $\Delta(1232)$ and higher resonances might usefully constrain these
two parameters.  In Figs.~\ref{fig:sigmadelta-p} and \ref{fig:sigmadelta-n}
we present calculations
of the unpolarized cross section and the spin observables
$\Sigma_3$, $\Sigma_{2x}$, and $\Sigma_{2z}$ in the energy
range 200--500 MeV.  
In these calculations, the parameter $a_{1,t}$ was adjusted by changing
the $\sigma$ mass in the asymptotic contribution
$A_1^{\rm as}(t) = A_1^{\rm as}(0)/(1-t/M_\sigma^2)$, where
$A_1^{\rm as}(0)$ was already fixed by
experimental data on $\alpha_E-\beta_M$.
The parameter $a_2$ was adjusted by adding to the asymptotic contribution
from the $\pi^0$-exchange a contribution of heavier exchanges,
i.e.\ by using in the dispersion relations the ansatz
$A_2^{\rm as}(t) = (A_2^{\pi^0}(t) + C) F(t)$, where $C$ was an adjusted
constant and $F(t)$ was a monopole form factor with the cut-off parameter
$\Lambda_\pi \simeq 700$ MeV.
In the unpolarized cross section for the proton, a change
of $\gamma_\pi$ from $-37$ to $-41$ is indistinguishable from a change
in the $\sigma$ mass from 500 to 700 MeV (which changes
$\alpha_{E2}-\beta_{M2}$ from 49 to 38).  However, these possibilities
are easily distinguished with $\Sigma_{2z}$, so that a combination
of unpolarized and polarized measurements in this energy range offers
the possibility of placing strong constraints on both 
$\gamma_\pi$ and $\alpha_{E2}-\beta_{M2}$.
Of course, any practical determination of the polarizabilites 
from Compton scattering data at energies of a few hundred MeV has to
take into account uncertainties in the photopion
multipoles used to evaluate the dispersion integrals. At the moment,
these uncertainties are not negligible (see Ref.~\cite{legs97a}).

\section{Summary}

The general structure of the Compton scattering amplitude from
the nucleon with polarized photons and/or polarized nucleons in
the initial and/or final state has been developed.  A low-energy
expansion of the amplitude to $\O(\omega^4)$ has been given in
terms of ten polarizabilities:  two dipole polarizabilities
$\alpha_E$ and $\beta_M$, two dispersion corrections to the dipole
polarizabilities $\alpha_{E\nu}$ and $\beta_{M\nu}$, two
quadrupole polarizabilities $\alpha_{E2}$ and $\beta_{M2}$, and
four spin polarizabilities $\gamma_{E1}$, $\gamma_{M1}$,
$\gamma_{E2}$, and $\gamma_{M2}$.  The physical significance of
the parameters has been discussed, and the relationship between
these and the cross section and spin observables below the pion
threshold has been established.  We have also presented
theoretical predictions of these parameters based both on
fixed-$t$ dispersion relations and chiral perturbation theory.
We have established that the range of validity of our expansion
extends to the pion threshold.  We have shown that low-energy
experiments will have to be very precise to resolve the
theoretical ambiguities in the polarizabilities.  However, we have
suggested that measurements at higher energy might help fix the most
theoretically uncertain of them, particulary the backward
spin polarizabilitiy $\gamma_\pi$ and the difference of
quadrupole polarizabilities $\alpha_{E2}-\beta_{M2}$.

\acknowledgments
A.L. appreciates the hospitality of the University of Illinois at
Urbana-Champaign and the Institut fur Kernphysik at Mainz where a part
of this work was carried out.  This work was supported in part by the
National Science Foundation under Grant No. 94-20787.

\appendix

\section{Multipole content of polarizabilities}

\subsection{Centre-of-mass amplitudes}

In the CM frame, the amplitude $T_{fi}$ of nucleon Compton scattering
can be represented by six functions $R_i$ of the energy $\omega$ and
the CM angle $\theta^*$ as \cite{lapi59,cont62}
\beq
   T_{fi} = 8\pi W \sum_{i=1}^6  \rho_i R_i(\omega,\theta^*),
\eeq
where $W=\sqrt s$ and the spin basis $\rho_i$ reads
\beqn
\label{basic-r_i}
  & \rho_1 =  \See, \quad  \rho_2 =  \Sss, \quad
    \rho_3 = i\Vee, \quad  \rho_4 = i\Vss, \nn
  & \rho_5 = i(\Sk\Sse - \kS\Ses), \quad
    \rho_6 = i(\kS\Sse - \Sk\Ses).
\eeqn
In the particular cases of forward or backward scattering,
\beqn
   \frac 1{8\pi W} \Big[ T_{fi} \Big]_{\theta=0} &=&
      \rho_1\,(R_1+R_2) + \rho_3\,(R_3+R_4+2R_5+2R_6), \nn
   \frac 1{8\pi W} \Big[ T_{fi} \Big]_{\theta=\pi} &=&
      \rho_1\,(R_1-R_2) + \rho_3\,(R_3-R_4-2R_5+2R_6).
\eeqn

Some authors \cite{hear62,ragu93,bern95,hols97b} utilize
a different spin
basis, and the following identities provide links
with the notation of those works:
\beqn
\label{r_i-links}
  & \Sek\Ske = x \rho_1 - \rho_2, \quad
   i\Vkk\See = x \rho_3 + \rho_4 - \rho_5, \nn
  &  i(\Vek\Ske - \Vke\Sek) = 2x \rho_3 - \rho_5, \nn
  &  i(\Ss\Sek - \sS\Ske)  =  2 \rho_3 - \rho_6,
\eeqn
where $x=\Skk$ and where we have used $\vec e \cdot \hat{\vec k}
 = \vec e^{\prime *} \cdot \hat{\vec k}' = 0$.%
\footnote{
A few other relations can be obtained from (\ref{r_i-links}) by doing a
dual transformation, $\vec e\to \vec s = \hat{\vec k}\times\vec e$,
$\vec s\to -\vec e = \hat{\vec k}\times\vec s$ (and the same for primed
vectors) which is just a $\pi/2$-rotation of the polarizations.
Under such a transformation, $\rho_1 \leftrightarrow \rho_2$, $\rho_3
\leftrightarrow \rho_4$, and $\rho_5 \leftrightarrow \rho_6$.}
In particular, the amplitudes $A_i$ from \cite{hols97b}
(we denote them here by $A_i^{\rm H}$) read
\beqn
  & A_1^{\rm H} = c\,(R_1+xR_2), \quad
    A_2^{\rm H} = -cR_2, \quad
    A_3^{\rm H} = c\,(R_3+xR_4+2xR_5+2R_6), \nn
  & A_4^{\rm H} = cR_4, \quad
    A_5^{\rm H} = -c\,(R_4+R_5), \quad
    A_6^{\rm H} = -cR_6,
\eeqn
where $c=4\pi W/m$.

The CM amplitudes $R_i$ are related to the invariant amplitudes
(\ref{defA}) by
\beqn
\label{RviaA}
 R_1 &=& C \Big\{ c_1 \Big({-}A_1-\frac{W^2}{m^2} A_3 \Big)
       -\frac{\nu}{m} c_2 A_5 -\frac{W}{m} c_3 A_6 \Big\}, \nn
 R_2 &=& C \Big\{ c_1 \Big( A_1-\frac{W^2}{m^2} A_3 \Big)
       +\frac{\nu}{m} c_2 A_5 -\frac{W}{m} c_3 A_6 \Big\}, \nn
 R_3 &=& C \Big\{
      (W-m)^2 \Big( (x-1)A_1 + (1+x)\frac{W^2}{m^2}A_3 \Big)
       -\frac{\nu}{m} c_3 A_5 -\frac{W}{m} c_2 A_6 \Big\} ,\nn
 R_4 &=& C \Big\{
      (W-m)^2 \Big( (1-x)A_1 + (1+x)\frac{W^2}{m^2}A_3 \Big)
       +\frac{\nu}{m} c_3 A_5 -\frac{W}{m} c_2 A_6 \Big\} ,\nn
 R_5 &=& C \Big\{
     (W-m)^2 \Big({-}A_1 - \frac{W^2}{m^2}A_3 \Big)
    +(W^2-m^2) \Big(A_2 + \frac{W^3}{m^3} A_4 \Big) \nn && \qquad{}
    + 2(W-m) \Big({-}\nu A_5 + \frac{W^2}{m} A_6 \Big) \Big\}, \nn
 R_6 &=& C \Big\{
     (W-m)^2 \Big(A_1 - \frac{W^2}{m^2}A_3 \Big)
    +(W^2-m^2) \Big({-}A_2 + \frac{W^3}{m^3} A_4 \Big) \nn && \qquad{}
    + 2(W-m) \Big(\nu A_5 + \frac{W^2}{m} A_6 \Big) \Big\}.
\eeqn
Here $x=\cos\theta^*$ and
\beqn
  &\dis C=\frac{(s-m^2)^2}{64\pi s^2}, \qquad
  c_1 = 4mW + (W-m)^2 (1-x), \nn &
  c_2 = 4W(W-m) - (W-m)^2 (1-x), \qquad
  c_3 = 4W^2 - (W-m)^2 (1-x).
\eeqn
Note also that the invariants $\nu$, $t$, $\eta$ are
\beq
   \nu = \frac{s-m^2+t/2}{2m},
     \quad t = \frac{(s-m^2)^2}{2s}(x-1),
     \quad \eta = \frac{(s-m^2)^2}{2m^2}(x+1).
\eeq
In the important case of low energies (or very heavy nucleon), one has
\beqn
\label{R-A-low-omega}
    R_1  = -c \, (A_3+A_6+A_1), && \quad
    R_2  = -c \, (A_3+A_6-A_1),    \nn
    R_3  = -c'\, (A_6 + A_5),   && \quad
    R_4  = -c'\, (A_6 - A_5),      \nn
    2R_5 =  c'\, (A_4+A_6+A_2), && \quad
    2R_6 =  c'\, (A_4+A_6-A_2)
\eeqn
up to higher orders in $\omega/m$. Here $c=\omega^2/4\pi$ and
$c'=\omega^3/4\pi m$.

A multipole expansion of the amplitudes $R_i$ has the form
\cite{ritu57,cont62,naga65}
\beqn
  R_1 &=& \sum_{l\ge 1} \{
        [(l+1)f_{EE}^{l+} + lf_{EE}^{l-}] (lP'_l+P''_{l-1})
      - [(l+1)f_{MM}^{l+} + lf_{MM}^{l-}] P''_l \}, \nn
  R_2 &=& \sum_{l\ge 1} \{
        [(l+1)f_{MM}^{l+} + lf_{MM}^{l-}] (lP'_l+P''_{l-1})
      - [(l+1)f_{EE}^{l+} + lf_{EE}^{l-}] P''_l \}, \nn
  R_3 &=& \sum_{l\ge 1} \{
        [f_{EE}^{l+} - f_{EE}^{l-}] (P''_{l-1} - l^2P'_l)
      - [f_{MM}^{l+} - f_{MM}^{l-}] P''_l
      + 2f_{EM}^{l+} P''_{l+1} - 2f_{ME}^{l+} P''_l \}, \nn
  R_4 &=& \sum_{l\ge 1} \{
        [f_{MM}^{l+} - f_{MM}^{l-}] (P''_{l-1} - l^2P'_l)
      - [f_{EE}^{l+} - f_{EE}^{l-}] P''_l
      + 2f_{ME}^{l+} P''_{l+1} - 2f_{EM}^{l+} P''_l \}, \nn
  R_5 &=& \sum_{l\ge 1} \{
        [f_{EE}^{l+} - f_{EE}^{l-}] (lP''_l + P'''_{l-1})
      - [f_{MM}^{l+} - f_{MM}^{l-}] P'''_l \nn && \quad{}
      + f_{EM}^{l+} [(3l+1)P''_l + 2P'''_{l-1}]
      - 2f_{ME}^{l+} [(l+1)P''_{l+1} + 2P'''_l] \}, \nn[2ex]
  R_6 &=& \sum_{l\ge 1} \{
        [f_{MM}^{l+} - f_{MM}^{l-}] (lP''_l + P'''_{l-1})
      - [f_{EE}^{l+} - f_{EE}^{l-}] P'''_l \nn && \quad{}
      + f_{ME}^{l+} [(3l+1)P''_l + 2P'''_{l-1}]
      - 2f_{EM}^{l+} [(l+1)P''_{l+1} + 2P'''_l] \}.
\eeqn
Here $P_l=P_l(x)$ are Legendre polynomials of $x=\cos\theta^*$.
The multipole amplitudes $f_{TT'}^{l\pm}$ with $T,T'=E,M$ correspond to
transitions $Tl\to T'l'$ and the superscript indicates the
angular momentum $l$ of the initial photon and the total angular momentum
$j=l\pm\frac12$. Due to T-invariance,
\beq
  f_{EM}^{l+} = f_{ME}^{(l+1)-}, \quad
  f_{ME}^{l+} = f_{EM}^{(l+1)-}.
\eeq
Keeping only dipole-dipole and dipole-quadrupole transitions in these
formulas, we obtain
\beqn
\label{R-low-multipoles}
  & R_1 = 2f_{EE}^{1+} + f_{EE}^{1-}, \quad
    R_2 = 2f_{MM}^{1+} + f_{MM}^{1-}, \nn
  & R_3 = f_{EE}^{1-} - f_{EE}^{1+} + 6f_{EM}^{1+}, \quad
    R_4 = f_{MM}^{1-} - f_{MM}^{1+} + 6f_{ME}^{1+}, \nn
  & R_5 = -6f_{ME}^{1+}, \quad
    R_6 = -6f_{EM}^{1+}
\eeqn
(plus higher multipoles, which introduce an angular dependence
to the amplitudes $R_i$).

\subsection{Polarizabilities to order $\O(\omega^3)$}

Low-energy expansions of the amplitudes $R_i$ are obtained from
(\ref{RviaA}) and (\ref{A-exp}) \cite{levc85}.  Leading nonvanishing
terms of $R_i$ are given by the Born term, in accordance with the
low-energy theorem by Gell-Mann-Goldberger-Low:
\beqn
  & \dis R_1^{\rm B} =  r_0 q^2 \Big( -1 + (1+x)\frac{\omega}{m} \Big)
    + \O(\omega^2), \quad
    R_2^{\rm B} = -r_0 q^2 \frac{\omega}{m} + \O(\omega^2), \nn
  & \dis R_3^{\rm B} = -r_0 q^2 \frac{\omega}{2m} + \O(\omega^2), \quad
    R_4^{\rm B} = -r_0 (\kappa+q)^2 \frac{\omega}{2m} + \O(\omega^2), \nn
  & \dis R_5^{\rm B} = \O(\omega^2), \quad
    R_6^{\rm B} = r_0 q (\kappa+q) \frac{\omega}{2m} + \O(\omega^2),
\eeqn
where $r_0=e^2/4\pi m$ and $eq$ is the electric charge of the nucleon.
Structure-dependent (i.e. non-Born) contributions to $R_i$ start with
the terms
\beqn
\label{R-vs-gammas}
    R_1^{\rm NB} =  \omega^2 \alpha_E + \O(\omega^3), && \quad
    R_2^{\rm NB} =  \omega^2  \beta_M + \O(\omega^3), \nn
    R_3^{\rm NB} =  \omega^3 (\gamma_1+2\gamma_3) + \O(\omega^4), && \quad
    R_4^{\rm NB} =  \omega^3  \gamma_2            + \O(\omega^4), \nn
    R_5^{\rm NB} = -\omega^3 (\gamma_2+ \gamma_4) + \O(\omega^4), && \quad
    R_6^{\rm NB} = -\omega^3  \gamma_3            + \O(\omega^4),
\eeqn
where coefficients are directly read out from Eq.\ (\ref{R-A-low-omega}):
\beqn
    \alpha_E = -\frac{1}{4\pi}  (a_3+a_6+a_1),   && \quad
     \beta_M = -\frac{1}{4\pi}  (a_3+a_6-a_1),      \nn
    \gamma_1 =  \frac{1}{4\pi m}(a_4-a_5-a_2),   && \quad
    \gamma_2 =  \frac{1}{4\pi m}(a_5-a_6),          \nn
    \gamma_3 =  \frac{1}{8\pi m}(a_2-a_4-a_6),   && \quad
    \gamma_4 =  \frac{1}{8\pi m}(a_6-a_2-a_4-2a_5).
\eeqn
Here the constants $\gamma_i$ are those that appeared in Eq.\
(\ref{gammas-def}).

The physical meaning of these polarizabilities can be understood by using
the multipole expansion (\ref{R-low-multipoles}).  Comparing with Eq.\
(\ref{R-vs-gammas}), we identify the
polarizabilities as leading terms in the structure-dependent
multipoles:
\beqn
  \omega^2\alpha_E &\simeq& (2f_{EE}^{1+} + f_{EE}^{1-})^{\rm NB}, \nn
  \omega^2 \beta_M &\simeq& (2f_{MM}^{1+} + f_{MM}^{1-})^{\rm NB}, \nn
  \omega^3\gamma_1  &\simeq& (f_{EE}^{1-} - f_{EE}^{1+}
         - 6f_{EM}^{1+})^{\rm NB}, \nn
  \omega^3\gamma_2  &\simeq& (f_{MM}^{1-} - f_{MM}^{1+}
         + 6f_{ME}^{1+})^{\rm NB}, \nn
  \omega^3\gamma_3  &\simeq& (6f_{EM}^{1+})^{\rm NB}, \nn
  \omega^3\gamma_4  &\simeq& (f_{MM}^{1+} - f_{MM}^{1-})^{\rm NB}.
\eeqn
It is seen that some of these $\gamma_i$ describe mixed effects of
spin-dependent dipole scattering and dipole-quadrupole transitions.
A more transparent physical meaning can be ascribed to the quantities
\beqn
  \gamma_{E1} = -\gamma_1 - \gamma_3
  = \frac{1}{8\pi m}(a_6 - a_4  + 2a_5 + a_2)
     &\simeq & \omega^{-3}(f_{EE}^{1+} - f_{EE}^{1-})^{\rm NB}, \nn
  \gamma_{M1} =  \gamma_4
  = \frac{1}{8\pi m}(a_6 - a_4  - 2a_5 - a_2)
     &\simeq & \omega^{-3}(f_{MM}^{1+} - f_{MM}^{1-})^{\rm NB},
\eeqn
which describe a spin dependence of the dipole transitions
$E1\to E1$ and $M1\to M1$, and to the quantities
\beqn
  \gamma_{E2} = \gamma_2 + \gamma_4
  = -\frac{1}{8\pi m}(a_4 + a_6  + a_2)
     &\simeq & 6\omega^{-3}(f_{ME}^{1+})^{\rm NB}, \nn
  \gamma_{M2} = \gamma_3
  = -\frac{1}{8\pi m}(a_4 + a_6  - a_2)
     &\simeq & 6\omega^{-3}(f_{EM}^{1+})^{\rm NB},
\eeqn
which describe transitions to quadrupole states, $M1\to E2$ and $E1\to 
M2$. In terms of these quantitities, the structure-dependent parts of 
the amplitudes $R_3$ to $R_6$ read
\beqn
\label{R-vs-alphas1}
  R_3^{\rm NB} &=& 
          \omega^3 ({-}\gamma_{E1} + \gamma_{M2}) + \O(\omega^4), \nn
  R_4^{\rm NB} &=& 
          \omega^3 ({-}\gamma_{M1} + \gamma_{E2}) + \O(\omega^4), \nn
  R_5^{\rm NB} &=& -\omega^3\gamma_{E2} + \O(\omega^4), \nn
  R_6^{\rm NB} &=& -\omega^3\gamma_{M2} + \O(\omega^4).
\eeqn
The quantities $\gamma_{E1}$, $\gamma_{M1}$, $\gamma_{E2}$, and
$\gamma_{M2}$ are related to
the spin polarizabilities $\delta$, $\kappa$, $\delta_1$ and $\kappa_1$
of Levchuk and Moroz \cite{levc85} (they call them
gyrations) by
\beq
  \textstyle
  \delta = -\gamma_{E1} + \gamma_{M2}, \quad
  \kappa = -\gamma_{M1} + \gamma_{E2}, \quad
  \delta_1 = -\gamma_{E2}, \quad
  \kappa_1 = -\gamma_{M2},
\eeq
to the spin polarizabilities $\gamma_i$ of Ragusa \cite{ragu93} by
\beq
  \textstyle
  \gamma_1= -\gamma_{E1} - \gamma_{M2}, \quad
  \gamma_2= -\gamma_{M1} + \gamma_{E2}, \quad
  \gamma_3=                \gamma_{M2}, \quad
  \gamma_4=  \gamma_{M1},
\eeq
to the spin polarizabilities $\alpha_1$, $\beta_1$, $\alpha_2$, $\beta_2$
of Babusci \ea \cite{babu97} by
\beq
  \textstyle
  \alpha_1=  4\gamma_{M2}, \quad
   \beta_1= -4\gamma_{E2}, \quad
  \alpha_2= -2\gamma_{E1} - 2\gamma_{M2}, \quad
   \beta_2=  2\gamma_{M1} + 2\gamma_{E2},
\eeq
and to the forward- and backward-angle spin polarizabilities by
\beq
\textstyle
  \gamma   = -\gamma_{E1} - \gamma_{M1} - \gamma_{E2} - \gamma_{M2},
\quad
  \gamma_\pi = -\gamma_{E1} + \gamma_{M1} + \gamma_{E2} - \gamma_{M2}.
\eeq
The parameter $\delta$ in \cite{babu97} is $\delta=-\gamma_\pi$.

\subsection{Quadrupole and dispersion polarizabilities}

Effects described by the constants $\alpha_\nu$, $\alpha_t$,
$\beta_\nu$, and $\beta_t$ correspond to the following contributions of
order $\O(\omega^4)$ to the spin-independent amplitudes $R_1$ and
$R_2$:
\beq
\label{R-vs-alphas2}
  \delta R_1^{(4)} = \omega^4 (\alpha_\nu + (2x-2)\alpha_t), \quad
  \delta R_2^{(4)} = \omega^4 ( \beta_\nu + (2x-2) \beta_t)
\eeq
(these are not all terms of order $\O(\omega^4)$ in $R_{1,2}$, as
is discussed in Appendix~C).
The constants in (\ref{R-vs-alphas2}) are related to
$\O(\omega^4)$ terms in dipole-dipole and quadrupole-quadrupole transitions.
Introducing weighed sums over projections of the total angular momentum $j$,
\beq
  f_{El} = (l+1)f_{EE}^{l+} + lf_{EE}^{l-}, \quad
  f_{Ml} = (l+1)f_{MM}^{l+} + lf_{MM}^{l-},
\eeq
we have
\beq
\label{R12}
  R_1 = f_{E1} + 2xf_{E2} -  f_{M2},  \quad
  R_2 = f_{M1} + 2xf_{M2} -  f_{E2}.
\eeq
Comparing with Eq.~(\ref{R-vs-alphas2}), we conclude that the constants
$\alpha_t$ and $\beta_t$ are proportional to the electric and magnetic
quadrupole polarizabilities of the nucleon \cite{rade79},
\beqn
\label{alpha-E2}
   \alpha_{E2} = 12\alpha_t
     & \simeq  & 12   \omega^{-4}(f_{E2})^{\rm NB}, \nn
    \beta_{E2} = 12 \beta_t
     & \simeq  & 12   \omega^{-4}(f_{M2})^{\rm NB}.
\eeqn
The normalization coefficient here is explained in Appendix~B.  It is
chosen to conform to the definitions used in atomic physics where,
for example, dynamic electric polarizabilities of the
hydrogen read \cite{rade79,au78}
\beq
  4\pi\alpha_{El}(\omega) = e^2 \sum_{n\ne 0}
  \Big( \frac1{E_n-E_0-\omega} + \frac1{E_n-E_0+\omega} \Big) \,
   | (r^l P_l(\cos\theta))_{n0} |^2 .
\eeq
The factor $4\pi$ arises because we use units in
which $e^2\simeq 4\pi/137$.

The combinations
\beqn
  \alpha_{E\nu} &=& \alpha_\nu - 2\alpha_t +  \beta_t, \nn
   \beta_{M\nu} &=&  \beta_\nu - 2 \beta_t + \alpha_t
\eeqn
are identified as $\O(\omega^4)$ terms in the dipole amplitudes $f_{E1}$
and $f_{M1}$; that is, dispersion effects in the dynamic dipole
polarizabilities
\beq
\label{aE,bM-disp}
  \alpha_{E1}(\omega) = \alpha_E + \omega^2\alpha_{E\nu} + \cdots, \qquad
   \beta_{M1}(\omega) =  \beta_M + \omega^2 \beta_{M\nu} + \cdots.
\eeq
For the hydrogen atom,
\beq
\label{alphaEv-sum}
  4\pi\alpha_{E} = 2 \sum_{n\ne 0}
   \frac{|(D_z)_{n0}|^2}{E_n-E_0},
\quad
  4\pi\alpha_{E\nu} = 2 \sum_{n\ne 0}
   \frac{|(D_z)_{n0}|^2}{(E_n-E_0)^3} , \quad \vec D=e\vec r.
\eeq

\section{Normalization of the polarizabilities}

In this appendix we explain the normalizations of the polarizabilities and
effective interactions, which are defined in Section~4, by using a
simple nonrelativistic model.  We discuss the quadrupole polarizability
$\alpha_{E2}$ and the spin polarizabilities $\gamma_{E2}$ and
$\gamma_{M1}$.

\subsection{Quadrupole polarizability $\alpha_{E2}$}

Let us consider a charged particle in the bound $s$-wave state
$|0\rangle$ affected by an external electric potential $A_0(\vec r)$.
Expanding the interaction $eA_0(\vec r)$ in powers of $\vec r$, we
get the quadrupole interaction with the external field,
\beq
\label{VE2}
  V = \frac12 e r_i r_j \nabla_i \nabla_j A_0(0)
         = -\frac16 Q_{ij} E_{ij}.
\eeq
Here $Q_{ij} = e\,(3r_i r_j - r^2 \delta_{ij})$ is the quadrupole
moment of the system and
\beq
   E_{ij}=-\nabla_i \nabla_j A_0 =
        \frac12 (\nabla_i E_j + \nabla_j E_i ), \quad  E_i^{~i}=0,
\eeq
is the quadrupole strength of the field.
The energy shift of the particle caused by the quadrupole interaction
(\ref{VE2}) is given by second-order perturbation theory,
\beq
\label{shiftE-E2}
  \Delta E = - \frac{1}{36} \sum_{n\ne 0}
     \frac{ (Q_{ij}E_{ij})_{0n} (Q_{pq}E_{pq})_{n0} } {E_n-E_0}
   = -\frac{1}{36} X E_{ij} E_{ij} .
\eeq
Here $n$ numerates excited states $|n\rangle$ and their energies $E_n$.
The quantity $X$ is defined as a coefficient in the expression
\beq
\label{sumQQ}
   \sum_{n\ne 0} \frac 1{E_n-E_0}
     \Big( (Q_{ij})_{0n} (Q_{pq})_{n0} + {\rm h.c.} \Big)
     = X (\delta_{ip}\delta_{jq} + \delta_{iq}\delta_{jp}
        -\frac23 \delta_{ij}\delta_{pq}) .
\eeq
The r.h.s. of Eq.~(\ref{sumQQ}) is the most general tensor $T_{ijpq}$
which has vanishing traces $T_{i.pq}^{~i} = T_{ijp.}^{~~~p}=0$ and
is symmetric under $i\leftrightarrow j$, $p\leftrightarrow q$, or
$ij\leftrightarrow pq$.

Taking $i,j,p,q=z$,
we relate $X$ to the quadrupole polarizability $\alpha_{E2}$:
\beq
\label{alphaE2-sum}
  X = 12\pi\alpha_{E2},
   \quad  4\pi\alpha_{E2}  \equiv \frac12 \sum_{n\ne 0}
  \frac { | (Q_{zz})_{n0} |^2 } {E_n-E_0} \, .
\eeq
Finally, the energy shift (\ref{shiftE-E2}) takes the form of an
effective quadrupole potential
\beq
   H_{\rm eff}^{E2\,\rm nospin} = - \frac{1}{12} \,
         4\pi\alpha_{E2} E_{ij} E_{ij}.
\eeq

\subsection{Quadrupole spin polarizability $\gamma_{E2}$}

Now let us consider a particle moving around a heavy nucleus. We assume
that both the particle and the nucleus have spin and that the total
spin of the system in the ground state $|0\rangle$ is 1/2.  In the
presence of both an electric quadrupole and magnetic dipole
interaction,
\beq
\label{VE2+M1}
  V = -\frac16 Q_{ij} E_{ij} - M_i H_i,
\eeq
where $M_i$ is the magnetic moment operator, the corresponding
energy shift of the system, $\sum_{n\ne 0} V_{0n}V_{n0}/(E_0-E_n)$,
has a mixed $E2$-$M1$ term,
\beq
\label{shiftE-E2-M1}
  \Delta E = -\frac16 \sum_{n\ne 0} \frac 1{E_n-E_0}
       \Big( (Q_{ij}E_{ij})_{0n} (M_k H_k)_{n0} + {\rm h.c.} \Big)
   = -\frac16 Y E_{ij}(\sigma_i H_j + \sigma_j H_i) .
\eeq
Here $Y$ is defined as
\beq
\label{sumQM}
   \sum_{n\ne 0} \frac 1{E_n-E_0}
         \Big( (Q_{ij})_{0n} (M_k)_{n0} + {\rm h.c.} \Big)
     = Y (\sigma_i\delta_{jk} + \sigma_j\delta_{ik}
        -\frac23 \sigma_k\delta_{ij}).
\eeq
The r.h.s. of Eq.~(\ref{sumQM}) is the most general tensor $T_{ijk}$
which has vanishing trace $T_{i.k}^{~i}=0$ and is symmetric under
$i\leftrightarrow j$.

Taking $i,j,k=z$, we relate $Y$ to the quadrupole spin polarizability
$\gamma_{E2}$:
\beq
\label{gammaE2-sum}
  Y = -12\pi\gamma_{E2},   \qquad
  4\pi\gamma_{E2}\sigma_z \equiv
    -\frac14 \sum_{n\ne 0} \frac{1}{E_n-E_0}
    \Big( (Q_{zz})_{0n} (M_z)_{n0} + {\rm h.c.} \Big),
\eeq
where the last equation explicitly shows the normalization and physical
meaning of $\gamma_{E2}$. Such a polarizability can exist if there are
tensor forces inside the system.
Finally, the energy shift (\ref{shiftE-E2}) takes the form of an
effective potential
\beq
   H_{\rm eff}^{E2,\,\rm spin} =  \frac12 \, 4\pi\gamma_{E2}
       E_{ij}(\sigma_i H_j + \sigma_j H_i)
       = 4\pi\gamma_{E2} E_{ij}\sigma_i H_j.
\eeq

\subsection{Dipole spin polarizability $\gamma_{M1}$}

Now, let us assume that the above spin-1/2 system scatters a photon
through a magnetic dipole interaction $-\vec M\cdot \vec H(t)$.  
Omitting the Born contribution, we write the Compton scattering amplitude
through intermediate excited states as
\beq
  \frac{1}{2m} T_{fi}^{M1}
   =  \omega^2\, s_j^{\prime *} s_i \sum_{n\ne 0} \Big(
    \frac{(M_j)_{0n} (M_i)_{n0}}{E_n-E_0-\omega} +
    \frac{(M_i)_{0n} (M_j)_{n0}}{E_n-E_0+\omega} \Big).
\eeq
Its spin dependent part at low energies is
\beq
\label{T-M1-spin-model}
  \frac{1}{8\pi m} T_{fi}^{M1,\,\rm spin} = -i\gamma_{M1} \omega^3 \Vss,
\eeq
where the parameter $\gamma_{M1}$ is defined as a coefficient in the
equation
\beq
\label{sumMM}
     \sum_{n\ne 0} \frac{1}{(E_n-E_0)^2}
    \Big( (M_j)_{0n} (M_i)_{n0} - {\rm h.c.} \Big) =
    4\pi\gamma_{M1} \, i\epsilon_{ijk} \sigma_k .
\eeq

The scattering amplitude (\ref{T-M1-spin-model}) can be associated with
an effective spin-dependent interaction
\beq
   H_{\rm eff}^{M1,\,\rm spin} = - \frac12 \, 4\pi\gamma_{M1} \,
      \vec\sigma \cdot \vec H \times \vec{\dot H}.
\eeq

\subsection{Compton scattering amplitude}

Now we give a summary of interactions and scattering amplitudes based 
on the above normalizations.  Note that the corresponding electric and 
magnetic effective interactions are related through the duality 
transformation, $E_i\to H_i$, $H_i\to -E_i$.

With the effective quadrupole interaction
\beq
   H_{\rm eff}^{(E2,M2),\,\rm nospin} = - \frac{1}{12} \, 4\pi
        (\alpha_{E2} E_{ij} E_{ij} + \beta_{M2} H_{ij} H_{ij}),
\eeq
where $H_{ij}= \frac12 (\nabla_i H_j + \nabla_j H_i)$ and $H_i^{~i}=0$,
the Compton scattering amplitude reads
\beq
  \frac{1}{8\pi m} T_{fi}^{(E2,M2),\,\rm nospin}
   =  \frac{1}{12}\omega^4 \alpha_{E2} (2z\rho_1 - \rho_2)
    + \frac{1}{12}\omega^4  \beta_{M2} (2z\rho_2 - \rho_1),
\eeq
where $\rho_i$ are given in (\ref{basic-r_i}). The spin-dependent
dipole-quadrupole interaction
\beq
   H_{\rm eff}^{(E2,M2),\,\rm spin} = 4\pi
     ( \gamma_{E2} E_{ij}\sigma_i H_j
      -\gamma_{M2} H_{ij}\sigma_i E_j )
\eeq
results in the Compton scattering amplitude
\beq
  \frac{1}{8\pi m} T_{fi}^{(E2,M2),\,\rm spin}
   =  \omega^3 \gamma_{E2} (\rho_4 - \rho_5)
    + \omega^3 \gamma_{M2} (\rho_3 - \rho_6).
\eeq
The spin-dependent dipole interaction
\beq
   H_{\rm eff}^{(E1,M1),\,\rm spin} = - \frac12 \, 4\pi
    \epsilon_{ijk}\sigma_k
     (\gamma_{E1} E_i\dot  E_j + \gamma_{M1} H_i\dot H_j )
\eeq
gives the Compton scattering amplitude
\beq
  \frac{1}{8\pi m} T_{fi}^{(E1,M1),\,\rm spin}
   = -\omega^3 (\gamma_{E1} \rho_3 + \gamma_{M1} \rho_4).
\eeq

\section{Quadrupole polarizabilities and \\
   relativistic corrections to the dipole interaction}

The polarizabilities of the nucleon can only be given
an exact meaning through definition.
The simplest definition of the multipole polarizabilities $\alpha_{El}$ and
$\beta_{Ml}$ is that they are the appropriately normalized
coefficients of the $\omega^{2l}$ terms in the partial-wave amplitudes
of Compton scattering, $(f_{El})^{\rm NB}$ and 
$(f_{Ml})^{\rm NB}$ \cite{rade79,guic95}.
However, we do not follow this approach for the
quadrupole polarizabilities because it leads to some unwanted features
when relativistic effects are taken into account. Considering
$\O(\omega^4)$ terms in the amplitudes, we want to exclude contributions
which are arise merely as
relativistic recoil corrections to the dipole polarizabilities.

We would like to associate with the polarizabilities $\alpha_{El}$ and
$\beta_{Ml}$ those nucleon-structure effects in the amplitude $T_{fi}$
which are even functions of the photon energy or momentum and do not
depend on the nucleon spin.  However, both the energy and the spin
depend on the reference frame. If the frame is changed, the energy
undergoes a Lorentz transformation and the Pauli spinors of the nucleon
undergo a Wigner rotation.  If the amplitude $T_{fi}$
associated with the polarizabilities $\alpha_{El}$, $\beta_{Ml}$ is
chosen to be spin independent in the CM frame, it would be spin
dependent in other frames, including the Lab and Breit frames.  Moreover,
since the electric and magnetic fields are not invariant under Lorentz
transformations, the splitting of structure effects into electric and
magnetic  parts, $\sim\alpha_{El}$ and $\sim\beta_{Ml}$, may also
depend on the frame.  Giving a relativistically sound definition, we
have to be cautious when choosing a frame and using
correspondence with notions of classical physics.

The CM frame is not good in this respect.  The CM
amplitudes $R_i$ do not possess all the symmetries which the amplitude
$T_{fi}$ itself has.  The crossing transformation,
\beq
  e,k \leftrightarrow e^{\prime *},-k',
\eeq
brings the total momentum of the $\gamma N$ system, $k+p$, out of rest,
so that the amplitudes $R_i$ are neither odd nor even
functions of the energy.  Therefore, an effective covariant interaction
({\it i.e.}, an effective Lagrangian), which describes the polarizabilities
and possesses the symmetries of the total amplitude $T_{fi}$,
would result in CM amplitudes $R_i$ which contain terms of
mixed order in $\omega$, both even and odd. It would be difficult to
rely on individual terms in $R_i$ when identifying the polarizabilities,
except for terms of lowest order.

The amplitudes in the Lab frame are also not good, because of the lack
of symmetry between the initial and final nucleon.  In particular,
the PT-transformation,
\beq
  e,k \leftrightarrow e',k', \qquad
  \vec\sigma,\vec p \leftrightarrow -\vec\sigma,\vec p',
\eeq
applied to the $\gamma N$ system, brings the initial nucleon out of
rest. That is why Eq.\ (\ref{Tlab}) contains both even and odd powers
of the photon energies.

The best choice is provided by the Breit frame, in which the nucleon
before and after photon scattering has the momentum
\beq
  \vec p_{\rm B} = \frac12(\vec k'-\vec k)_{\rm B}
    =\vec Q_{\rm B}, \quad \vec p'_{\rm B}=-\vec Q_{\rm B},
\eeq
respectively.  In such a frame, both T-invariance and crossing
symmetry are fulfilled in the simplest way and, importantly,
the nucleon is at rest on average, $(\vec p+\vec p')_{\rm B}=0$.
That is why, in the course of an analysis of elastic $eN$-scattering,
the Breit frame rather than the CM frame is used to relate the
amplitude of the reaction $eN\to eN$ with physically meaningful
structure functions of the nucleon, the electromagnetic form factors
$G_E$ and $G_M$.  For some deeper motivation in favor of the Breit
frame and its relation with the language of wave packets, see Ref.\
\cite{sach60}.

Therefore in constructing our definitions, we choose to postulate that the
polarizability interaction and the related Compton scattering amplitude
are spin independent in the Breit frame.  It will be
spin dependent in the CM frame.  Since the nucleon spin in the Lab
or Breit frame is the same,%
\footnote{That is because the Wigner angles for the nucleon-spin
rotation between the Lab and Breit frames vanish for both the
initial and final nucleon.  The Wigner angle, $\vec\theta_W \propto
\vec V\times \vec v$, depends on the velocity $\vec v$ of the nucleon
itself and the relative velocity $\vec V$ of the frames. In the case of
the transformation between the Lab and Breit frames, $\vec\theta_W=0$
for the initial nucleon $N$, because it is at rest, $\vec v=\vec p=0$.
Also, $\vec\theta_W=0$ for the final nucleon $N'$, because $\vec
V\propto \vec p'+\vec p=\vec p'$ is parallel to the nucleon velocity
$\vec v\propto \vec p'$.  The similar Wigner angles for photons are
generally not zero, so that the photon polarizations are different in
the Lab and Breit frames.}
the amplitude in the Lab frame will be spin independent also.
Nevertheless, we do not wish to directly relate individual
spin-independent terms in Eq.\ (\ref{Tlab}) of different orders in
$\omega\omega'$ with appropriate polarizabilities.  That is partly
because the amplitude (\ref{Tlab}) is not symmetric with respect to the
initial and final nucleon.  The factors $\omega\omega'\See$ and
$\omega\omega'\Sss$ in Eq.\ (\ref{Tlab}) represent the electric and
magnetic fields taken in the rest frame of the initial nucleon, whereas
a sound definition should use the fields in the frame in which the
nucleon is at rest, at least on average.  Some of the
$\O(\omega^2\omega^{\prime 2})$ terms in Eq.\ (\ref{Tlab})
are actually the result of a Lorentz transformation of
$\O(\omega^2)$ terms in the Breit frame.

For the above reasons, we choose as a definition of the Compton
scattering amplitude related with the dipole polarizabilities the
expression
\beq
\label{TBreit(alpha,beta)}
   T_{fi,\,\rm B}^{(\alpha_E,\,\beta_M)} \equiv
          4\pi\omega^2_{\rm B} \bar u' u\,
  (\vec e^{\prime *}_{\rm B}\cdot \vec e_{\rm B}\,\alpha_E
 + \vec s^{\prime *}_{\rm B}\cdot \vec s_{\rm B}\,\beta_M),
\eeq
where both the energy $\omega_{\rm B}$ and all polarizations are taken
in the Breit frame. Neither $\O(\omega^4)$ terms
nor recoil corrections
$\sim t/m^2$ are explicitly included here. The factor $\bar u'
u=\sqrt{4m^2-t} = 2mN(t)$ is spin independent in the Breit frame and
serves only for a covariant normalization.%
\footnote{{\it Cf.}\ the definition of the electromagnetic form factors of
the nucleon \cite{sach60}.}
Note also that $\omega_{\rm B}=\nu /N(t)$.  Since the spin-independent
part of the Compton scattering amplitude in the Breit frame reads
\beqn
\label{TBreit}
  T_{fi,\,\rm B}^{\rm nospin} &=&
     \bar u'(p')\, e^{\prime * \mu} \Big\{ {-}
   \frac{P'_\mu P'_\nu}{P'^2} \Big(T_1 + \frac{m^2\nu T_2}{m^2-t/4}\Big)
  -\frac{N_\mu N_\nu}{N^2} \Big(T_3 + \frac{m^2\nu T_4}{m^2-t/4}\Big)
     \Big\} e^\nu u   \nn &=&
  \frac{2m\nu^2}{N(t)} \Big\{
     \See\Big[ {-}A_1 - \Big(1-\frac{t}{4m^2}\Big) A_3
         - \frac{\nu^2 A_5}{m^2-t/4} - A_6 \Big]
 \nn && \qquad {} +
     \Sss\Big[    A_1 - \Big(1-\frac{t}{4m^2}\Big) A_3
         + \frac{\nu^2 A_5}{m^2-t/4} - A_6 \Big] \Big\},
\eeqn
the invariant amplitudes $A_i$ corresponding to
Eq.~(\ref{TBreit(alpha,beta)}) are
\beq
  A_1^{(\alpha_E,\beta_M)} = -2\pi(\alpha_E-\beta_M), \quad
  A_3^{(\alpha_E,\beta_M)} = -\frac{2\pi(\alpha_E+\beta_M)}{1-t/4m^2},
\eeq
and other $A_i^{(\alpha_E,\beta_M)}$ are zero.
Using Eq.~(\ref{Tlab}), we find the corresponding scattering
amplitude in the Lab frame:
\beq
\label{Tlab(alpha,beta)}
   T_{fi}^{(\alpha_E,\,\beta_M)} =
          \frac{8\pi m \omega\omega'}{N(t)} \Big\{
  (\See\alpha_E + \Sss\beta_M
  -\frac{t}{4m^2}(\See-\Sss)(\alpha_E-\beta_M) \Big\},
\eeq
where a recoil correction $\sim t/m^2$ appears as a result of the
no-recoil {\it ansatz} in the Breit frame, Eq.~(\ref{TBreit(alpha,beta)}).

With the above definition of the contribution of the dipole
polarizabilities, we write the remaining terms of the non-Born amplitude
$T_{fi}^{\rm NB,\,nospin}$ as
\beq
   T_{fi}^{\rm NB,\,nospin} - T_{fi}^{(\alpha_E,\,\beta_M)} =
     8\pi m \omega\omega' \Big\{
   \See(\nu^2\alpha_\nu + t\alpha_t)
  +\Sss(\nu^2 \beta_\nu + t \beta_t) \Big\} + \O(\omega^6).
\eeq
They are given by the parameters $\alpha_\nu$, $\alpha_t$, $\beta_\nu$
and $\beta_t$ in Eq.~(\ref{alpha-nu-t-def}), which determine quadrupole
and dispersion polarizabilities, as discussed in Appendix~A.

\section{Pole contribution of the $\Delta(1232)$ to polarizabilities}

To calculate the contribution of the $\Delta$-isobar excitation into
the polarizabilities, we write an effective $\gamma N\Delta$
interaction in the form similar to Eq.~(\ref{VE2+M1}):
\beq
\label{Heff-Delta}
   H_{\rm eff} = -\vec M\cdot\vec H - \frac16 Q_{ij}\nabla_i E_j.
\eeq
Here $\vec M$ and $Q_{ij}$ are the magnetic dipole and electric
quadrupole transition operators and are characterized by the matrix elements
\beq
  \langle \Delta, \textstyle \pm\frac12 |
   M_z | N, \textstyle \pm\frac12 \rangle = \mu_{N\Delta}, \qquad
  \langle \Delta, \textstyle \pm\frac12 |
   Q_{zz} | N, \textstyle \pm\frac12 \rangle = \pm Q_{N\Delta}.
\eeq
Since the interaction (\ref{Heff-Delta}) involves $M1$ and $E2$
transitions into the $j=3/2$ state, it contributes to the multipoles
$f_{MM}^{1+}$, $f_{EE}^{2-}$, $f_{ME}^{1+}$ and  therefore to the
polarizabilities $\beta_M$, $\gamma_{M1}$, $\alpha_{E2}$, $\gamma_{E2}$.

Using these matrix elements and Eq.~(\ref{alphaE2-sum})
from Appendix~B, we find
$4\pi\alpha_{E2}= Q_{N\Delta}^2/(2\Delta)$.
From Eq.~(\ref{gammaE2-sum}) we get
$4\pi\gamma_{E2}= -\mu_{N\Delta} Q_{N\Delta} /(2\Delta)$.
Using the same matrix elements, the Wigner-Eckart theorem,
and Eq.~(\ref{sumMM}), we find
$4\pi\gamma_{M1}= \mu_{N\Delta}^2 /\Delta^2$.
Finally, the equations
$4\pi\beta_M = 2 \mu_{N\Delta}^2 / \Delta$
and
$4\pi\beta_{M\nu} = 2 \mu_{N\Delta}^2 / \Delta^3$
are magnetic analogs of Eq.~(\ref{alphaEv-sum}).

In terms of the ratio $R = E_{1+}/M_{1+}$ of the resonance multipoles
of pion photoproduction taken at the resonance energy ($E_{\gamma,\,\rm
lab}=340$ MeV),
\beq
   \frac{Q_{N\Delta}}{\mu_{N\Delta}} =
       \frac{12}{k} R \simeq -0.25~{\rm fm},
\eeq
where $k$ is the photon energy of the decay $\Delta\to\gamma N$ in the
rest frame of the $\Delta$ and $R \simeq -2.75\%$ (we take an average
of $-2.5\%$ \cite{beck97} and $-3.0\%$ \cite{legs97}).

\newpage

\end{document}